\documentclass[a4paper,12pt]{article}
\pdfoutput=1
\usepackage{amsmath}
\usepackage{mathrsfs}
\usepackage{dsfont}
\usepackage{bbm,bm}
\usepackage{graphicx,subfigure}
\usepackage{xcolor}
\usepackage[colorlinks,
            linkcolor=blue,
            filecolor=blue,
            anchorcolor=blue,
            urlcolor=blue,
            citecolor=blue,
            bookmarks=false,
            ]{hyperref}
\usepackage[numbers,sort&compress]{natbib}            
\usepackage{url}
\usepackage{setspace}
\usepackage{array}
\usepackage{booktabs}

\newlength{\dinwidth}
\newlength{\dinmargin}
\makeatletter
\newcommand{\thickhline}{%
    \noalign {\ifnum 0=`}\fi \hrule height 1pt
    \futurelet \reserved@a \@xhline
}
\makeatother

\allowdisplaybreaks

\def \SM{{\rm SM}}
\def \NP{{\rm NP}}
\def \TeV{{\,\rm TeV}}

\def \GeV{{\,\rm GeV}}

\def \ab{{\,\rm ab}}

\newcommand{\AFB}{A_{\rm FB}}

\newcommand{\mC}{{\mathcal C}}
\newcommand{\mB}{{\mathcal B}}

\def \la{\lambda}
\def \lb{\Lambda_b}
\def \lc{\Lambda_c}
\def\Dst{{D^*}}
\def\nn{\nonumber}

\def\bra#1{\left\langle #1\right|}
\def\ket#1{\left| #1\right\rangle}

\newenvironment{spmatrix}{\Biggl(\begin{matrix}} {\end{matrix}\Biggr)}

\newcommand{\abs}[1]{\left\vert{#1}\right\vert}
\begin{document}

\setlength{\abovedisplayskip}{6pt}
\setlength{\belowdisplayskip}{6pt}

\title{\bf Phenomenology of \boldmath{$b\to c\tau\bar\nu$} decays in a scalar leptoquark model}

\author{
  Han Yan\footnote{yanhan@mails.ccnu.edu.cn},\,
  Ya-Dong Yang\footnote{yangyd@mail.ccnu.edu.cn},\,
  and\,
  Xing-Bo Yuan\footnote{y@mail.ccnu.edu.cn}\\[15pt]
\small Institute of Particle Physics and Key Laboratory of Quark and Lepton Physics~(MOE), \\[-0.2cm]
\small Central China Normal University, Wuhan, Hubei 430079, China}

\date{}

\maketitle
\vspace{0.2cm}

\begin{abstract}
  \noindent During the past few years, hints of Lepton Flavour Universality (LFU) violation have been observed in $b \to c \tau \bar\nu$ and $b \to s \ell^+ \ell^-$ transitions. Recently, the $D^*$ and $\tau$ polarization fractions $P_L^{D^*}$ and $P_L^\tau$ in $B \to D^* \tau \bar\nu$ decay have also been measured by the Belle collaboration. Motivated by these intriguing results, we revisit the $R_{D^{(*)}}$ and $R_{K^{(*)}}$ anomalies in a scalar leptoquark (LQ) model, in which two scalar LQs, one being $SU(2)_L$ singlet and the other $SU(2)_L$ triplet, are introduced simultaneously. We consider five $b \to c \tau \bar\nu$ mediated decays, $B \to D^{(*)}\tau \bar\nu$, $B_c \to \eta_c \tau \bar\nu$, $B_c \to J/\psi \tau  \bar\nu$, and $\Lambda_b \to \Lambda_c \tau \bar\nu$, and focus on the LQ effects on the $q^2$ distributions of the branching fractions, the LFU ratios, and the various angular observables in these decays. Under the combined constraints by the available data on $R_{D^{(*)}}$, $R_{J/\psi}$, $P_L^\tau(D^*)$, and $P_L^{D^*}$, we perform scans for the LQ couplings and make predictions for a number of observables. It is found numerically that both the differential branching fractions and the LFU ratios are largely enhanced by the LQ effects, with the latter being expected to provide testable signatures at the SuperKEKB and High-Luminosity LHC experiments.\\[1.5em]
\end{abstract}

\newpage

\section{Introduction}
\label{sec:intro}

So far, the LHC has not observed any direct evidence for New Physics (NP) particles beyond the Standard Model (SM). However, several hints of Lepton Flavour University (LFU) violation emerge in the measurements of semileptonic $b$-hadron decays, which, if confirmed with more precise experimental data and theoretical predictions, would be unambiguous signs of NP~\cite{Li:2018lxi,Bifani:2018zmi}.

The charged-current decays $B \to D^{(*)} l \bar\nu$, with $\ell=e$, $\mu$ or $\tau$, have been measured by the BaBar~\cite{Lees:2012xj,Lees:2013uzd}, Belle~\cite{Huschle:2015rga,Sato:2016svk,Hirose:2016wfn,Hirose:2017dxl} and LHCb~\cite{Aaij:2015yra,Aaij:2017uff,Aaij:2017deq} collaborations. For the ratios of the branching fractions\footnote{Compared to the branching fractions themselves, the ratios $R_{D^{(\ast)}}$ are advantaged by the fact that, apart from significant reduction of the experimental systematic uncertainties, the CKM matrix element $V_{cb}$ cancels out and the sensitivity to $B\to D^{(\ast)}$ transition form factors becomes much weaker.}, $R_{D^{(*)}} \equiv \mathcal B(B\to D^{(*)}\tau\bar\nu) / \mathcal B(B\to D^{(*)}\ell\bar\nu)$, with $\ell=e$ and/or $\mu$, the latest experimental averages by the Heavy Flavor Averaging Group read~\cite{HFLAV:2016}
\begin{align}
R_D^{\rm exp}&=0.407\pm 0.039 \, ({\rm stat.}) \pm 0.024 \, ({\rm syst.}),  \\
R_{D^*}^{\rm exp}&=0.306\pm 0.013 \, ({\rm stat.})\pm 0.007 \, ({\rm syst.}),\, 
\end{align}
which exceed their respective SM predictions~\cite{HFLAV:2016}\footnote{Here the SM values are the arithmetic averages~\cite{HFLAV:2016} of the most recent calculations by several groups~\cite{Bigi:2016mdz,Jaiswal:2017rve,Bernlochner:2017jka,Bigi:2017jbd}.}
\begin{align}
  R_D^\SM&=0.299\pm 0.003, \qquad R_{D^{\ast}}^\SM =0.258\pm 0.005,
\end{align}
by $2.3\sigma$ and $3.0\sigma$, respectively. Considering the experimental correlation of $-0.203$ between $R_D$ and $R_{D^*}$, the combined results show about $3.78\sigma$ deviation from the SM predictions~\cite{HFLAV:2016}. Such a discrepancy, referred to as the $R_{D^{(*)}}$ anomaly, may provide a hint of LFU violating NP~\cite{Li:2018lxi,Bifani:2018zmi}. For the $B_c \to J/\psi \ell \bar\nu$ decay, a ratio $R_{J/\psi}$ can be defined similarly, and the recent LHCb measurement, $R_{J/\psi}^{\rm exp}=0.71\pm 0.17\,({\rm stat.}) \pm 0.18\, ({\rm syst.})$~\cite{Aaij:2017tyk}, lies about $2\sigma$ above the SM prediction, $R_{J/\psi}^\SM=0.248 \pm 0.006$~\cite{Wang:2008xt}. In addition, the LHCb measurements of the ratios $R_{K^{(*)}}\equiv \mB(B \to K^{(*)} \mu^+ \mu^-)/\mB(B \to K^{(*)} e^+ e^-)$, $R_K^{\rm exp}=0.745_{-0.074}^{+0.090}\pm0.036$ for $1.0\GeV^2\leq q^2\leq 6.0\GeV^2$~\cite{Aaij:2014ora} and $R_{K^*}^{\rm exp}=0.69_{-0.07}^{+0.11}\pm 0.05$ for $1.1\GeV^2\leq q^2\leq 6.0\GeV^2$~\cite{Aaij:2017vbb}, are found to be about $2.6\sigma$ and $2.5\sigma$ lower than the SM expectation, $R_{K^{(*)}}^{\rm SM}\simeq1$~\cite{Hiller:2003js,Bordone:2016gaq}, respectively.  These anomalies have motivated numerous studies both in the Effective Field Theory approach~\cite{Sakaki:2014sea,Bhattacharya:2014wla,Calibbi:2015kma,Feruglio:2016gvd,Choudhury:2017qyt,Hu:2018veh} and in specific NP models~\cite{Celis:2012dk,Deshpand:2016cpw,Celis:2016azn,Wei:2018vmk,Li:2018rax,Hu:2018lmk}. We refer to refs.~\cite{Li:2018lxi,Bifani:2018zmi} for recent reviews.

Recently, the Belle collaboration reported the first preliminary result of the $D^*$ longitudinal polarization fraction in the $B \to D^* \tau \bar\nu$ decay~\cite{Adamczyk:2019wyt,Abdesselam:2019wbt}
\begin{align}
  P_L^{D^*}=0.60 \pm 0.08 \, (\text{stat.}) \pm 0.04 \, (\text{syst.}),
\end{align}
which is consistent with the SM prediction $P_{L}^{D^*}=0.46 \pm 0.04$~\cite{Alok:2016qyh} at $1.5\sigma$. Together with the measurements of the $\tau$ polarization, $P_L^\tau =-0.38 \pm 0.51 \, (\text{stat.})_{-0.16}^{+0.21} \, (\text{syst.})$~\cite{Hirose:2016wfn,Hirose:2017dxl}, they provide valuable information about the spin structure of the interaction involved in $B \to D^{(*)} \tau \bar\nu$ decays, and are good observables to test various NP scenarios~\cite{Alok:2016qyh,Huang:2018nnq,Aebischer:2018iyb,Blanke:2018yud,Iguro:2018vqb,Tanaka:2012nw}. Measurements of the angular observables in these decays will be considerably improved in the future~\cite{Kou:2018nap,Bediaga:2018lhg}. For example, the Belle II experiment with $50\ab^{-1}$ data can measure $P_L^\tau$ with an expected precision of $\pm 0.07$~\cite{Kou:2018nap}.

In this work, motivated by these experimental progresses and future prospects, we study five $b \to c \tau \bar\nu$ decays, $B \to D^{(*)}\tau \bar\nu$, $B_c \to \eta_c \tau \bar\nu$, $B_c \to J/\psi \tau  \bar\nu$, and $\Lambda_b \to \Lambda_c \tau \bar\nu$, in the leptoquark (LQ) model proposed in ref.~\cite{Crivellin:2017zlb}. Models with one or more LQ states, which are colored bosons and couple to both quarks and leptons, are one of the most popular scenarios to explain the $R_{D^{(*)}}$ and $R_{K^{(*)}}$ anomalies~\cite{Dorsner:2016wpm,Freytsis:2015qca,Bauer:2015knc,Li:2016pdv,Fajfer:2015ycq,Deppisch:2016qqd,Dumont:2016xpj,DiLuzio:2017vat,Cai:2017wry,Calibbi:2017qbu,Kumar:2018kmr,Hati:2018fzc,Crivellin:2018yvo,Angelescu:2018tyl,Kim:2018oih,Marzocca:2018wcf,Buttazzo:2017ixm,Arnan:2019olv}. In ref.~\cite{Crivellin:2017zlb}, the SM is extended with two scalar LQs, one being $SU(2)_L$ singlet and the other $SU(2)_L$ triplet. The model is also featured by the fact that these two LQs have the same mass and hypercharge and their couplings to fermions are related via a discrete symmetry. In this way, the anomalies in $b \to c \tau \bar\nu$ and $b \to s \mu^+ \mu^-$ transitions can be explained simultaneously, while avoiding potentially dangerous contributions to $b \to s \nu \bar\nu$ decays. By taking into account the recent developments on the transition form factors~~\cite{Wang:2008xt,Detmold:2015aaa,Bigi:2016mdz,Jaiswal:2017rve,Datta:2017aue,FLAG}, we derive constraints on the LQ couplings in this model. Then, predictions in the LQ model are made for the five $b \to c \tau \bar\nu$ decays, focusing on the $q^2$ distributions of the branching fractions, the LFU ratios, and the various angular observables. Implications for future searches at the High-Luminosity LHC (HL-LHC)~\cite{Cerri:2018ypt} and SuperKEKB~\cite{Kou:2018nap} are also briefly discussed.

This paper is organized as follows: In section~\ref{sec:model}, we give a brief review of the LQ model proposed in ref.~\cite{Crivellin:2017zlb}. In section~\ref{sec:framework}, we recapitulate the theoretical formulae for the various flavour processes, and discuss the LQ effects on these decays. In section~\ref{sec:numerical results}, we present our detailed numerical analysis and discussions. Our conclusions are given in section~\ref{sec:conclusions}. The relevant transition form factors and helicity amplitudes are presented in the appendices.

\section{The Model}
\label{sec:model}

In this section, we recapitulate the LQ model proposed in ref.~\cite{Crivellin:2017zlb}, where a scalar LQ singlet $\Phi_1$ and a triplet $\Phi_3$ are added to the SM field content, to explain the observed flavour anomalies. Under the SM gauge group $\big(\text{SU(3)}_C,\, \text{SU(2)}_L,\,\text{U(1)}_Y \big)$, the LQ states $\Phi_1$ and $\Phi_3$ transform as $(\boldsymbol{3},\boldsymbol{1},\boldsymbol{-2/3})$ and $(\boldsymbol{3},\boldsymbol{\bar{3}},\boldsymbol{-2/3})$, respectively. Their interactions with the SM fermions are described by the Lagrangian~\cite{Crivellin:2017zlb}  
\begin{align}\label{eq:lag}
\mathcal L=\lambda_{jk}^{1L}\bar{Q}_j^ci\tau_2 L_k \Phi_1^\dag +\lambda_{jk}^{3L}\bar{Q}_j^c i\tau_2(\tau\cdot\Phi_3)^\dag L_k+ \text{h.c.},
\end{align}
where $Q_j$ and $L_k$ denote the left-handed quark and lepton doublet with generation indices $j$ and $k$, respectively. The couplings $\lambda_{jk}^{1L}$ and $\lambda_{jk}^{3L}$ are complex in general, but taken to be real throughout this work. It is further assumed that these two scalar LQs have the same mass $M$, and their couplings to the SM fermions satisfy the following discrete symmetry~\cite{Crivellin:2017zlb}:
\begin{align}\label{eq:symmetry}
  \lambda_{jk}^L\equiv \lambda_{jk}^{1L},
  \qquad
  \lambda_{jk}^{3L} =e^{i\pi j} \lambda_{jk}^L.
\end{align}
With these two assumptions, the tree-level LQ contributions to the $b \to s \nu \bar\nu$ decays are canceled. After rotating to the mass eigenstate basis, the LQ couplings to the left-handed quarks involve the CKM elements as
\begin{align}
  \lambda_{d_j k}^L = \lambda_{jk}^{L},
  \qquad
  \lambda_{u_j k}^L=V_{ji}^* \lambda_{ik}^{L},
\end{align}
where $V_{ij}$ is the CKM matrix element.

\section{Theoretical Framework}
\label{sec:framework}

In this section, we shall introduce the theoretical framework for the relevant flavour processes, and discuss the LQ effects on these decays.

\subsection{$\boldsymbol{b \to c \tau \bar\nu}$ mediated processes}

Including the LQ contributions, the effective Hamiltonian responsible for $b \to c \ell_i \bar\nu_j$ transitions is given by~\cite{Crivellin:2017zlb}
\begin{align}\label{eq:Heff}
\mathcal H_{\rm{eff}} = \frac{4 G_{F}}{\sqrt 2}{V_{cb}} \mC_L^{ij}\bigl( \bar c \gamma ^\mu P_L b  \bigr) \bigl( \bar \ell_i \gamma _\mu P_L \nu _j \bigr),
\end{align}
with the Wilson coefficient $\mC_L^{ij} = \mC_L^{\SM, ij} + \mC_L^{\NP, ij}$. The $W$-exchange contribution within the SM gives $\mC_L^{\SM,ij}=\delta_{ij}$, and the LQ contributions result in
\begin{align}\label{eq:LQ:b2c}
	\mC_L^{{\rm NP},ij} = \frac{\sqrt 2 }{8 G_{F} M^2}\frac{V_{ck}}{V_{cb}}\lambda _{3j}^L\lambda _{ki}^{L*} \left[ 1 + (-1)^k\right].
\end{align}
It is noted that this Wilson coefficient is given at the matching scale $\mu_{\rm NP} \sim M$. However, as the corresponding current is conserved, we can obtain the low-energy Wilson coefficient, $\mC_L^{\NP, ij}(\mu_b)=\mC_L^{\NP, ij}$, without considering the Renormalization Group Evolution~(RGE) effect.

In this work, we consider five processes mediated by the quark-level $b \to c \ell \bar\nu$ transition, including $B \to D^{(*)} \ell \bar\nu$, $B_c \to \eta_c \ell \bar\nu$, $B_c \to J/\psi \ell \bar\nu$, and $\Lambda_b \to \Lambda _c \ell \bar\nu$ decays. All these processes can be uniformly represented by
\begin{align}\label{eq:decayproc}
M(p_M,\lambda_M)\to N(p_N,\lambda_N)+\ell^{-}(p_\ell,\lambda_\ell)+\bar\nu_\ell(p_{\bar\nu_\ell}),
\end{align}
where $(M,N)=(B, D^{(*)}),\,(B_c,\eta_c)\,,(B_c, J/\psi)$, and $(\Lambda_b, \Lambda_c)$, and $(\ell ,\bar\nu)=(e,\bar\nu_e),\,(\mu, \bar\nu_\mu)$, and $(\tau, \bar\nu_\tau)$. For each particle $i$ in the above decay, its momentum and helicity are denoted by $p_i$ and $\lambda_i$, respectively. In particular, the helicity of a pseudoscalar meson is zero, i.e., $\lambda_{B_{(c)},D,\eta_c}=0$. After averaging over the non-zero helicity of the hadron $M$, the differential decay rate of this process can be written as~\cite{Hagiwara:1989cu,Tanaka:2012nw}
\begin{align}\label{eq:dga}
\text{d}\Gamma^{\lambda_N,\,\lambda_\ell}(M \to N \ell^-\bar\nu_\ell)=\frac{1}{2m_M}\frac{1}{2|\lambda_{M}|+1}\sum_{\lambda_{M}}|\mathcal{M}^{\lambda_{M}}_{\lambda_{N},\lambda_\ell}|^2\text{d}\Phi_3,
\end{align}
with the phase space 
\begin{align}
  \text{d}\Phi_3 =\frac{\sqrt{Q_+Q_-}}{256\pi^3 m_{M}^2}\sqrt{1-\frac{m_{\ell}^2}{q^2}}\text{d}q^2\text{d}\cos \theta_\ell ,
\end{align}
where $Q_\pm =m_{\pm}^2 - q^2 $, with $m_\pm = m_{M} \pm m_{N}$ and $q^2$ the dilepton invariant mass squared. $\theta_\ell\in [0,\pi]$ denotes the angle between the three-momentum of $\ell$ and that of $N$ in the $\ell$-$\bar\nu$ center-of-mass frame. The helicity amplitudes $\mathcal M_{\lambda_N,\lambda_\ell}^{\lambda_M} \equiv \langle N\ell \bar{\nu}_\ell |\mathcal{H}_{\rm eff}|M\rangle$ can be written as~\cite{Datta:2017aue}
\begin{align}\label{eq:amp}
  \mathcal{M}_{\lambda_{N},\lambda_\tau}^{\lambda_{M}}=\frac{G_F V_{cb}}{\sqrt{2}} \Big( \!H^{SP}_{\lambda_{M},\lambda_{N}}L^{SP}_{\lambda_\tau} &+\sum_{\lambda_W}\eta_{\lambda_W}H^{VA}_{{\lambda_{M},\lambda_{N},\lambda_W}}L^{VA}_{\lambda_\tau,\lambda_W}\nn\\
  &+\sum_{\lambda_{W_1},\lambda_{W_2}} \eta_{\lambda_{W_1}} \eta_{\lambda_{W_2}} H^{T}_{\lambda_{M},\lambda_{N},\lambda_{W_1},\lambda_{W_2}}L^{T}_{\lambda_\tau,\lambda_{W_1}\lambda_{W_2}} \Big),
\end{align}
where $\lambda_{W_i}$ denotes the helicity of the virtual vector bosons $W$, $W_1$ and $W_2$. The coefficient $\eta_{\lambda_{W_i}}=1$ for $\lambda_{\lambda_{W_i}}=t$, and $\eta_{\lambda_{W_i}}=-1$ for $\lambda_{\lambda_{W_i}}=0,\,\pm 1$. Explicit analytical expressions of the leptonic and hadronic helicity amplitudes $H$ and $L$ are given in appendices \ref{sec:helicity amplitude} and \ref{sec:obs}. 

Starting with eq.~(\ref{eq:dga}), we can derive the following observables:
\begin{itemize}
\item The differential decay width and branching fraction
  \begin{align}
    \frac{{\rm d}\mathcal B(M\to N \ell\bar\nu_\ell)}{{\rm d}q^2}&=\frac{1}{\Gamma_M}\frac{{\rm d}\Gamma(M\to N \ell\bar\nu_\ell)}{{\rm d}q^2}\,\nn\\
    &=\frac{1}{\Gamma_M}\sum_{\lambda_N,\lambda_\ell}\frac{{\rm d}\Gamma^{\lambda_N,\lambda_\ell}(M\to N \ell\bar\nu_\ell)}{{\rm d}q^2}
  \end{align}
  where $\Gamma_{M}=1/\tau_{M}$ is the total width of the hadron $M$.
  
\item The $q^2$-dependent LFU ratio
  \begin{align}\label{eq:RLc}
  R_{N}(q^2)=\frac{{\rm d}\Gamma(M\to N \tau\bar\nu_\tau)/{\rm d}q^2}{{\rm d}\Gamma(M\to N l\bar\nu_l)/{\rm d}q^2}\,,
  \end{align}
  where $\text{d}\Gamma(M\to N l \bar\nu_l)/\text{d}q^2$ denotes the average of the different decay widths of the electronic and muonic modes.
  
\item The lepton forward-backward asymmetry
  \begin{align}\label{eq:AFB}
  A_{\rm FB}(q^2) = \frac{\int_{0}^{1} {\rm d}\cos\theta_\ell\,({\rm d}^2\Gamma/{\rm d}q^2{\rm d}\cos\theta_\ell)-\int_{-1}^{0}{\rm d}\cos\theta_\ell\,({\rm d}^2\Gamma/{\rm d}q^2{\rm d} \cos\theta_\ell )}{{\rm d}\Gamma/{\rm d}q^2}.
  \end{align}

\item The $q^2$-dependent polarization fractions
  \begin{align}\label{eq:PL}
     P_L^{\tau}(q^2)&=\frac{{\rm d}\Gamma^{\lambda_{\tau}=+1/2}/{\rm d}q^2-
                      {\rm d}\Gamma^{\lambda_{\tau}=-1/2}/{\rm d}q^2}{{\rm d}\Gamma/{\rm d}q^2}, && 
    \\
    P_L^N(q^2)&=\frac{{\rm d}\Gamma^{\lambda_N=+1/2}/{\rm d}q^2-
                {\rm d}\Gamma^{\lambda_N=-1/2}/{\rm d}q^2}{{\rm d}\Gamma/{\rm d}q^2}, \qquad &&\text{for } N=\Lambda_c,\nn
    \\
    P_L^N(q^2)&=\frac{{\rm d}\Gamma^{\lambda_N=0}/{\rm d}q^2}{{\rm d}\Gamma/{\rm d}q^2}, \qquad &&\text{for } N=D^*, J/\psi,\nn
    \\                                                                          P_T^{N}(q^2)&=\frac{{\rm d}\Gamma^{\lambda_N=1}/{\rm d}q^2-{\rm d}\Gamma^{\lambda_N=-1}/{\rm d}q^2}{{\rm d}\Gamma/{\rm d}q^2}, \qquad &&\text{for } N=D^*, J/\psi.\nn
\end{align}
\end{itemize}
Analytical expressions of all the above observables are given in appendix~\ref{sec:obs}. As these angular observables are ratios of the decay widths, they are largely free of hadronic uncertainties, and thus provide excellent tests of the NP effects.

As shown in eq.~(\ref{eq:Heff}), the LQ effects generate an operator with the same chirality structure as in the SM. Therefore, it is straightforward to derive the following relation:
\begin{align}\label{eq:RDr}
\frac{R_N}{R_N^\SM}= \sum\limits_{i = 1}^3 {{{\left\lvert {{\delta _{3i}} + \mC_L^{3i}} \right\rvert}^2}}\,,
\end{align}
with $N=D^{(*)}, \eta_c, J/\psi$, and $\Lambda_c$. Here, vanishing contributions to the electronic and muonic channels are already assumed.

One of the main inputs in our calculations are the transition form factors. In this respect, notable progresses have been achieved in recent years~\cite{Bigi:2016mdz,Jaiswal:2017rve,Bernlochner:2017jka,Bigi:2017jbd,Bernlochner:2019ldg,Wang:2017jow,Gubernari:2018wyi,Murphy:2018sqg,Berns:2018vpl,Wang:2018duy,Leljak:2019eyw,Gutsche:2015mxa,Detmold:2015aaa,Datta:2017aue,FLAG}. In this work, we adopt the Boyd-Grinstein-Lebed (BGL)~\cite{Boyd:1997kz,Bigi:2016mdz} and the Caprini-Lellouch-Neubert (CLN)~\cite{Caprini:1997mu,Jaiswal:2017rve} parametrization for the $B\to D$ and $B\to D^{*}$ transition form factors, respectively. In these approaches, both the transition form factors and the CKM matrix element $|V_{cb}|$ are extracted from the experimental data simultaneously. In addition, we use the $B_c\to \eta_c,\,J/\psi$ transition form factors obtained in the covariant light-front approach~\cite{Wang:2008xt}. For the $\Lambda_b \to \Lambda_c$ transition form factor, we adopt the recent Lattice QCD results in refs.~\cite{Detmold:2015aaa,Datta:2017aue}. Explicit expressions of all the relevant transition form factors are recapitulated in appendix~\ref{sec:form factor}.

\subsection{Other processes}

With the LQ effects considered, the effective Hamiltonian for $b \to s \ell_i^+ \ell_j^-$ transition can be written as~\cite{Buchalla:1995vs}
\begin{align}
  \mathcal H_{\rm eff}=- \frac{ 4 G_{F} }{\sqrt 2}V_{tb}V_{ts}^{*} \sum_a \mC_a^{ij} \mathcal O_a^{ij} + \text{h.c.},
\end{align}
where the operators relevant to our study are
\begin{align}
  \mathcal O_9^{ij} =\frac{\alpha_e }{4\pi} \bigl( \bar s \gamma ^\mu  P_L b \bigr) \bigl(\bar\ell_i \gamma _\mu \ell_j \bigr),
  \qquad
  \mathcal O_{10}^{ij} =\frac{\alpha_e }{4\pi} \bigl( \bar s \gamma ^\mu P_L b \bigr) \bigl(\bar\ell_i \gamma _\mu  \gamma^5\ell_j \bigr).
\end{align}
The LQ contributions result in~\cite{Crivellin:2017zlb} 
\begin{align}
\mC_9^{\NP,ij} =  - \mC_{10}^{\NP,ij} = \frac{-\sqrt 2 }{2 G_{F} V_{tb}V_{ts}^*} \frac{\pi }{\alpha_e }\frac{1}{M^2}\lambda _{3j}^{L}\lambda _{2i}^{L*}.
\end{align}
In the model-independent approach, the current $b\to s \mu^+ \mu^-$ anomalies can be explained by a $\mC_{9}^{\NP,22} =  - \mC_{10}^{\NP,22}$ like contribution, with the allowed range given by~\cite{Altmannshofer:2015sma,Descotes-Genon:2015uva,Hurth:2016fbr}
\begin{align}
-0.91 \, (-0.71) \leq \mC_{9}^{\NP,22}=-\mC_{10}^{\NP,22} \leq  -0.18 \, (-0.35)\,,
\end{align} 
at the $2\sigma$~($1\sigma$) level, which provides in turn a constraint on $\lambda_{22}^{L*}\lambda_{32}^L$. Furthermore, the LQ contributions to $b \to s \tau^+ \tau^-$ and $b \to c \tau \bar\nu_\tau$ transitions depend on the same product $\lambda_{23}^{L*}\lambda_{33}^L$, making therefore a direct correlation between the branching fraction $\mB(B_s \to \tau^+ \tau^-)$ and $R_{D^{(*)}}$.

For the $b \to s \nu \bar\nu$ transitions, both the LQs $\Phi_1$ and $\Phi_3$ generate  tree-level contributions. However, after assuming that they have the same mass, their effects are canceled out due to the discrete symmetry in eq.~(\ref{eq:symmetry}). In addition, this LQ scenario can accommodate the $(g-2)_\mu$ anomaly~\cite{Bennett:2006fi,PDG:2018}, once the right-handed interaction term $\lambda_{fi}^R\bar{u}_f^c\ell_i \Phi_1^\dagger$ is introduced to eq.~(\ref{eq:lag})~\cite{Crivellin:2017zlb}. We do not consider such a term in this work. More details can be found in ref.~\cite{Crivellin:2017zlb}, in which various lepton flavour violating decays of leptons and $B$ meson have also been discussed.

Finally, we give brief comments on direct searches for the LQs at high-energy colliders. Since the LQ contributions to $b \to c \tau \bar\nu$ transitions only involve the product $\lambda_{23}^{L*}\lambda_{33}^L$, searches for the LQs with couplings to the second and third generations are more relevant to our work. At the LHC, both the CMS and ATLAS collaborations have performed searches for such LQs in several channels, e.g., ${\rm LQ} \to t \mu$~\cite{Sirunyan:2018ruf}, ${\rm LQ}\to t \tau$~\cite{Aaboud:2019bye}, ${\rm LQ} \to b \tau$~\cite{Aaboud:2019bye}, etc. Current results from the LHC have excluded the LQs with masses below about $1\TeV$~\cite{PDG:2018}. For example, searches for pair-produced scalar LQs decaying into $t$ quark and $\mu$ lepton have been performed by the CMS Collaboration, in which a scalar LQ with mass below $1420\GeV$ have been excluded at $95\%$ CL with the assumption of $\mathcal B ({\rm LQ} \to t \mu)=1$~\cite{Sirunyan:2018ruf}. It is noted that all these collider constraints depend on the assumption of the total width of the LQ, which involve all the LQ couplings $\lambda_{ij}^L$. In order to apply the collider constraints to our scenario, one need to perform a global fit on all the LQ couplings and derive bounds on the total width. Such analysis is out of the scope of this paper. For the scenario with one singlet and one triplet LQ, we refer to ref.~\cite{Marzocca:2018wcf} for more detailed collider analysis. In addition, it is noted that our analysis does not depend on the mass of the LQ, since the LQ couplings always appear in the form of $\lambda_{23}^{L*}\lambda_{33}^L/M^2$ in $b \to c \tau \bar\nu$ transitions as in eq.~(\ref{eq:LQ:b2c}).

%Therefore, the bounds in eq.~(\ref{eq:lam2333}) are equivalent to $ -2.90 \, [\TeV^{-2}] <\lambda _{23}^{L*}\lambda _{33}^L/M^2 <-2.74 \, [\TeV^{-2}]$ and $ 0.03 \, [\TeV^{-2}] <\lambda _{23}^{L*}\lambda _{33}^L/M^2 < 0.20 \, [\TeV^{-2}]$.

%LQs have accordingly been searched for and studied in the context of $e^+e^-$~\cite{Doncheski:1996un,Doncheski:1996dp},~$ep$~\cite{Plehn:1997az,Friberg:1997nn},~$p\bar{p}$~\cite{Abazov:2011qj,Aaltonen:2007rb} and~$pp$~\cite{Sirunyan:2018nkj,Sirunyan:2018btu,Aaboud:2019jcc} colliders. Using data corresponding to ~$35.9\fb^{-1}$ at $13\TeV$, the CMS collaboration has performed search for heavy scalar LQs in $pp\to t\bar{t}\tau^+\tau^-$ channel. The mass of LQs is excluded up to $900\GeV$ at $95\%$ CL~\cite{Sirunyan:2018nkj}. CMS~\cite{Sirunyan:2018btu} finds that 1st generation LQs with masses below $1.45(1.27)$\TeV~are excluded for $\mB(S_{LQ}\to l^{\pm}j)=1(0.5)$. ATLAS~\cite{Aaboud:2019jcc} gets a similar result that lower limits on LQs mass at $95\%$ CL extend up to $1.25$\TeV~for 1st and 2nd generation LQs. The LQs mass limits are expected to become even more stringent in the near future with the LHC run II and HL-LHC data~\cite{Dorsner:2016wpm}.

\section{Numerical Analysis}
\label{sec:numerical results}

In this section, we proceed to present our numerical analysis of the LQ effects on the decays considered. After deriving the constraints on the model parameters, we concentrate on its effects on the five $b\to c \tau \bar\nu$ decays, i.e., $B \to D^{(*)}\tau \bar\nu$, $B_c \to \eta_c \tau\bar\nu$, $B_c \to J/\psi \tau \bar\nu$, and $\Lambda_b \to \Lambda_c \tau\bar\nu$.

\subsection{SM predictions}

\begin{table}[t]
  \tabcolsep 0.12in
  \setlength{\extrarowheight}{5pt}
  \begin{center}
  \caption{\label{tab:inputs} Input parameters used in our numerical analysis.}
  \vspace{0.18cm}
  \begin{tabular}{cccc}
    \toprule
    \toprule
    Input & Value & Unit & Ref.\\
    \midrule
    $\alpha_s(m_Z)$ & $0.1181 \pm 0.0011$ & &  \cite{PDG:2018}
    \\
    $m_t^{\rm pole}$ & $173.1\pm 0.9$ &${\rm GeV}$ & \cite{PDG:2018}
    \\
    $m_b(m_b)$ & $4.18\pm0.03$ &${\rm GeV}$& \cite{PDG:2018}
    \\
    $m_c(m_c)$ & $1.275\pm0.025$ &${\rm GeV}$& \cite{PDG:2018}
    \\
    $|V_{cb}|$(semi-leptonic) & $41.00\pm 0.33\pm 0.74$ &$10^{-3}$ &\cite{Charles:2004jd}
    \\
    $|V_{ub}|$(semi-leptonic) & \hphantom{0}$3.98\pm 0.08\pm 0.22$ &$10^{-3}$ &\cite{Charles:2004jd}
    \\
    \bottomrule
    \bottomrule
  \end{tabular}
  \end{center}
\end{table}

\begin{table}[t]
	\tabcolsep 0.10in   
	\setlength{\extrarowheight}{5pt}  
	\begin{center}
	\caption{\label{tab:numerical results} Predictions for the branching fractions~(in unit of $10^{-2}$) and the ratios $R_N$ of the five $b \to c \tau \bar\nu $ decay modes in the SM and the LQ scenario. The entry ``\rule{3em}{1pt}'' indicates that no measurement is yet available for the corresponding observable.}
	\vspace{0.18cm} 
	\begin{tabular}{ccccc}
		\toprule
		\toprule
		Observable& SM & NP &Exp & Ref\\
		\midrule
		$ \mB (B\rightarrow D\tau\bar\nu) $ & $0.711_{-0.041}^{+0.042}$ & $[0.702,0.991]$ &$0.90\pm0.24$ &\cite{PDG:2018}
		\\
		$R_D$ & $0.301_{-0.003}^{+0.003}$ &$[0.313,0.400]$ & $0.407\pm0.039\pm0.024$ & \cite{HFLAV:2016}
		\\
		$ \mB(B_c \to \eta_c\tau\bar\nu) $&$0.204_{-0.024}^{+0.024}$ & $[0.188,0.299]$& \rule{3em}{1pt} &
		\\
		$R_{\eta_c}$ &  $0.281_{-0.031}^{+0.035}$ & $[0.263,0.416]$&\rule{3em}{1pt} &
		\\
		$ \mB(B\to D^{*}\tau\bar\nu) $ & $1.261_{-0.085}^{+0.087}$ & $[1.234,1.788]$ &$1.78\pm0.16$ &\cite{PDG:2018}
		\\
		$R_{D^*}$ & $0.258\pm 0.008$ & $[0.263,0.351]$ & $0.306\pm0.013\pm0.007$ & \cite{HFLAV:2016}
		\\
		$P_L^\tau$ & $-0.503\pm0.013$\hphantom{$-$} & $[-0.516,-0.490]$ &$-0.38\pm0.51_{-0.16}^{+0.21}$ & \cite{Hirose:2016wfn,Hirose:2017dxl}
		\\
		$P_{L}^{D^{*}}$ & $0.453\pm0.012$ & $[0.441,0.465]$ &$0.60 \pm 0.08 \pm 0.04$ & \cite{Adamczyk:2019wyt,Abdesselam:2019wbt}
		\\
		$ \mB(B_c\to J/\psi\tau\bar\nu) $& $0.398_{-0.049}^{+0.045}$ &$[0.366,0.583]$
		&\rule{3em}{1pt} &
		\\
		$R_{J/\psi}$ & $0.248_{-0.005}^{+0.006}$ & $[0.255,0.335]$&$0.71\pm0.17\pm0.18$ & \cite{Aaij:2017tyk}
		\\
		$\mB(\Lambda_b\to \Lambda_c\tau\bar\nu)$ &  $1.762_{-0.104}^{+0.105}$ &$[1.737,2.457]$ &\rule{3em}{1pt} &
		\\
		$R_{\lc}$ &$0.333_{-0.010}^{+0.010}$ &  $[0.339,0.451]$
		&\rule{3em}{1pt} &
		\\
		\bottomrule
		\bottomrule
	\end{tabular}
    \end{center}
\end{table}

In table~\ref{tab:inputs}, we collect the relevant input parameters used in our numerical analysis. Using the theoretical framework described in section~\ref{sec:framework}, the SM predictions for $B \to D^{(*)}\tau \bar\nu$, $B_c \to \eta_c \tau\bar\nu$, $B_c \to J/\psi \tau \bar\nu$, and $\Lambda_b \to \Lambda_c \tau\bar\nu$ decays are given in table~\ref{tab:numerical results}. To obtain the theoretical uncertainties, we vary each input parameters within their respective $1\sigma$ range and add each individual uncertainty in quadrature. For the uncertainties of transition form factors, the correlations among the fit parameters have been taken into account. In particular, for the $\Lambda_b \to \Lambda_c \tau \bar\nu$ decay, we follow the treatment of ref.~\cite{Detmold:2015aaa} to obtain the statistical and systematic uncertainties induced by the $\Lambda_b \to \Lambda_c$ transition form factors. From table~\ref{tab:numerical results}, it is found that the experimental data on the ratios $R_D$, $R_{D^*}$, and $R_{J/\psi}$ deviate from the SM predictions by $2.31\sigma$, $2.85\sigma$ and $1.83\sigma$, respectively.

\subsection{Constraints}

\begin{figure}[t]
	\centering
	\includegraphics[width=0.45\textwidth]{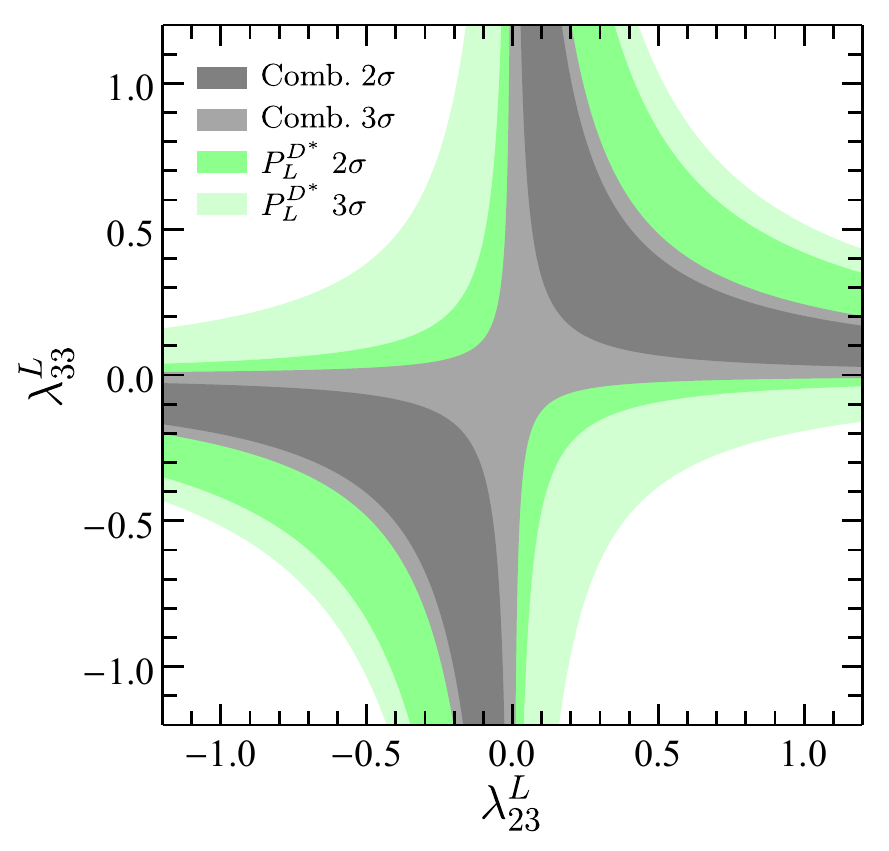}
	\caption{Combined constraints on $(\lambda_{23}^L, \lambda_{33}^L)$ by all the $b \to c \tau \bar\nu$ processes at $2\sigma$ (black) and $3\sigma$ (gray) levels. The dark (light) green area indicates the allowed region by $P_L^{D^*}$ only at $2\sigma$ ($3\sigma$).}
	\label{fig:lam23lam33}
\end{figure}

To get the allowed ranges of the LQ parameters, we impose the experimental constraints in the same way as in refs.~\cite{Jung:2012vu,Chiang:2017etj}; i.e., for each point in the parameter space, if the difference between the corresponding theoretical prediction and experimental data is less than $2\sigma$ ($3\sigma$) error bar, which is calculated by adding the theoretical and experimental uncertainties in quadrature, this point is regarded as allowed at $2\sigma$ ($3\sigma$) level.

In the LQ scenario introduced in section~\ref{sec:model}, the LQ contributions to $b \to c \tau \bar\nu$ transitions are all controlled by the product $\lambda_{23}^{L*}\lambda_{33}^L$. In the following analysis, the couplings $\lambda_{23}^L$ and $\lambda_{33}^L$ are assumed to be real. After considering the current experimental measurements of $R_{D^{(*)}}$, $R_{J/\psi}$, $P_L^\tau(D^*)$, and $P_L^{D^*}$, we find that constraints on $\lambda_{23}^{L*}\lambda_{33}^L$ are dominated by $R_D$ and $R_{D^*}$. The allowed ranges of $\lambda_{23}^{L*}\lambda_{33}^L$ at $2\sigma$ level are obtained to be as
\begin{align}\label{eq:lam2333}
 -2.90<\lambda _{23}^{L*}\lambda _{33}^L <-2.74, \qquad \text{or} \qquad 0.03<\lambda _{23}^{L*}\lambda _{33}^L<0.20,
\end{align}
where a common LQ mass $M=1\TeV$ is taken. The solution with negative $\lambda_{23}^{L*}\lambda_{33}^L$ corresponds to the case in which the LQ interactions dominate over the SM contributions. We do not pursue this possibility in the following analysis. For the solution with positive $\lambda_{23}^{L*}\lambda_{33}^L$, the allowed regions of $(\lambda_{23}^L,\lambda_{33}^L)$ at both $2\sigma$ and $3\sigma$ levels are shown in figure~\ref{fig:lam23lam33}. In this figure, we also show the individual constraint from the $D^*$ polarization fraction $P_L^{D^*}$, which is still weaker than the ones from $R_{D^{(*)}}$. In addition, the current measurement of the $\tau$ polarization fraction $P_L^\tau$ in $B\to D^* \tau \nu$ decay cannot give any relevant constraint.

\begin{figure}[t]
	\centering
	\includegraphics[width=0.45\linewidth]{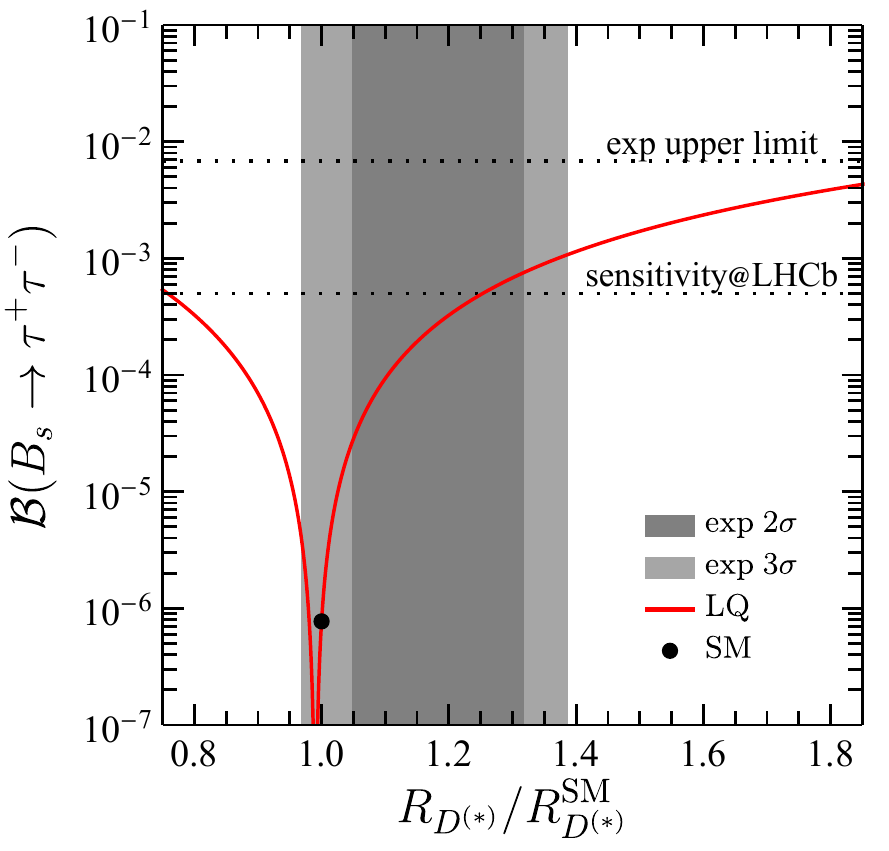}
	\caption{Correlation between $R_{D^{(*)}}/R_{D^{(*)}}^\SM$ and $\mB(B_s\to\tau^+ \tau^- )$. The black (gray) region denotes the $2\sigma$ ($3\sigma$) experimental ranges of $R_{D^{(*)}}/R_{D^{(*)}}^\SM$. The horizontal dashed and dotted lines correspond to the current LHCb upper limit and the expected sensitivity by the end of LHCb Upgrade II, respectively. The black point indicates the SM predictions.}
	\label{fig:correlation}
\end{figure}

As mentioned already in section~\ref{sec:framework}, the LQ contributions to $b \to s \tau^+ \tau^-$ and $b \to c \tau \bar\nu_\tau$ depend on the same product $\lambda_{23}^{L*}\lambda_{33}^L$. In the case of positive $\lambda_{23}^{L*}\lambda_{33}^L$, we show in figure~\ref{fig:correlation} the correlation between $R_{D^{(*)}}/R_{D^{(*)}}^\SM$ and  $\mB (B_s \to \tau^+ \tau^-)$. It can be seen that the LQ effects enhance the branching fraction of $B_s \to \tau^+ \tau^-$ in most of the parameter space. At present, the experimental upper limit $6.8 \times 10^{-3}$~\cite{Aaij:2017xqt} is far above the SM prediction $( 7.73 \pm 0.49) \times 10^{-7}$~\cite{Bobeth:2013uxa}. However, in order to produce the $2\sigma$ experimental range of $R_{D^{(*)}}$, the LQ contributions enhance $\mB (B_s \to \tau^+ \tau^-) $ by about $2$-$3$ orders of magnitude compared to the SM prediction, which reaches the expected LHCb sensitivity $5 \times 10^{-4}$ by the end of Upgrade II~\cite{Albrecht:2017odf,Bediaga:2018lhg}. It is noted that the $B \to K^{(*)} \tau^+ \tau^-$ decay may also play an important role in probing the LQ effects. Although the Belle II experiment will improve the current upper limit $2.25 \times 10^{-3}$ at $90\%$ confidence level by no more than two orders of magnitude, the proposed FCC-$ee$ collider can provide a few thousand of $B^0 \to K^{*0} \tau^+ \tau^-$ events from $\mathcal O (10^{13})$ $Z$ decays~\cite{Kamenik:2017ghi}.

\subsection{Predictions}

Using the constrained parameter space at $2\sigma$ level derived in the last subsection, we make predictions for the five $b \to c \tau \bar\nu$ processes. Table.~\ref{tab:numerical results} shows the SM and LQ predictions for the branching fractions $\mathcal B$ and LFU ratios $R$ of $B \to D^{(*)}\tau \bar\nu$, $B_c \to \eta_c \tau\bar\nu$, $B_c \to J/\psi \tau \bar\nu$, and $\Lambda_b \to \Lambda_c \tau\bar\nu$ decays. The LQ predictions have included the uncertainties induced by the transition form factors and CKM matrix elements. Considering that the polarization fractions $P_L^\tau$ and $P_L^{D^*}$ have already been measured, their SM and LQ predictions are also shown in table~\ref{tab:numerical results}. It can be seen that, although the LQ predictions for the branching fractions $\mB$ and the LFU ratios $R$ of the $B_c \to \eta_c \tau \bar\nu$ and $B_c \to J/\psi \tau \bar\nu$ decays lie within the $1\sigma$ range of their respective SM values, they can be significantly enhanced by the LQ effects.

\begin{figure}[t]
  \begin{center}
    \includegraphics[width=0.39\textwidth]{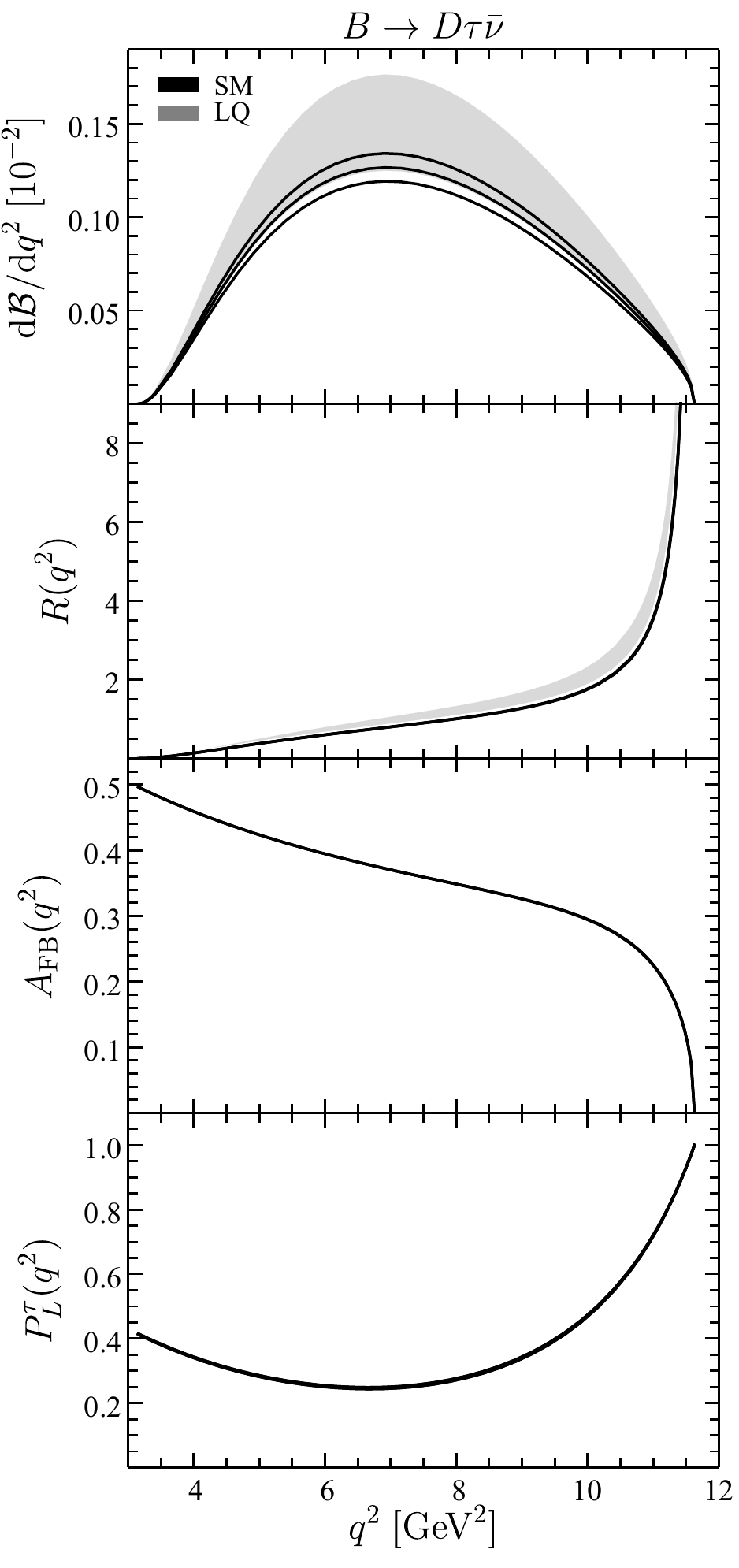}
    \qquad
    \includegraphics[width=0.39\textwidth]{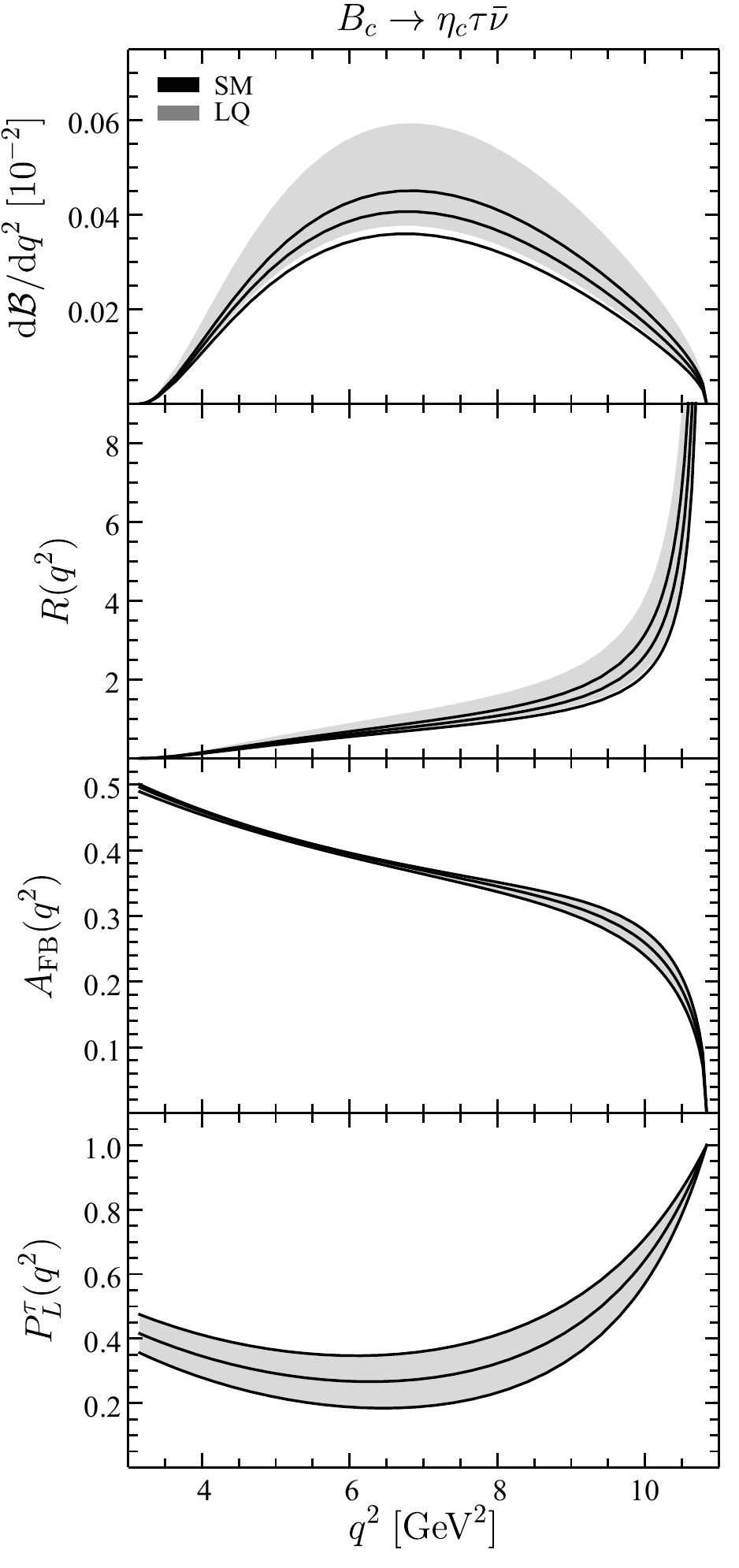}
    \caption{The $q^2$ distributions of the observables in $B \to D \tau \bar\nu$ (left) and $B_c \to \eta_c \tau \bar\nu$ (right) decays. The black curves (gray band) indicate the SM (LQ) central values with $1\sigma$ theoretical uncertainty. }
    \label{fig:B2PS}
  \end{center}
\end{figure}

\begin{figure}[t]
  \begin{center}
    \includegraphics[width=0.39\linewidth]{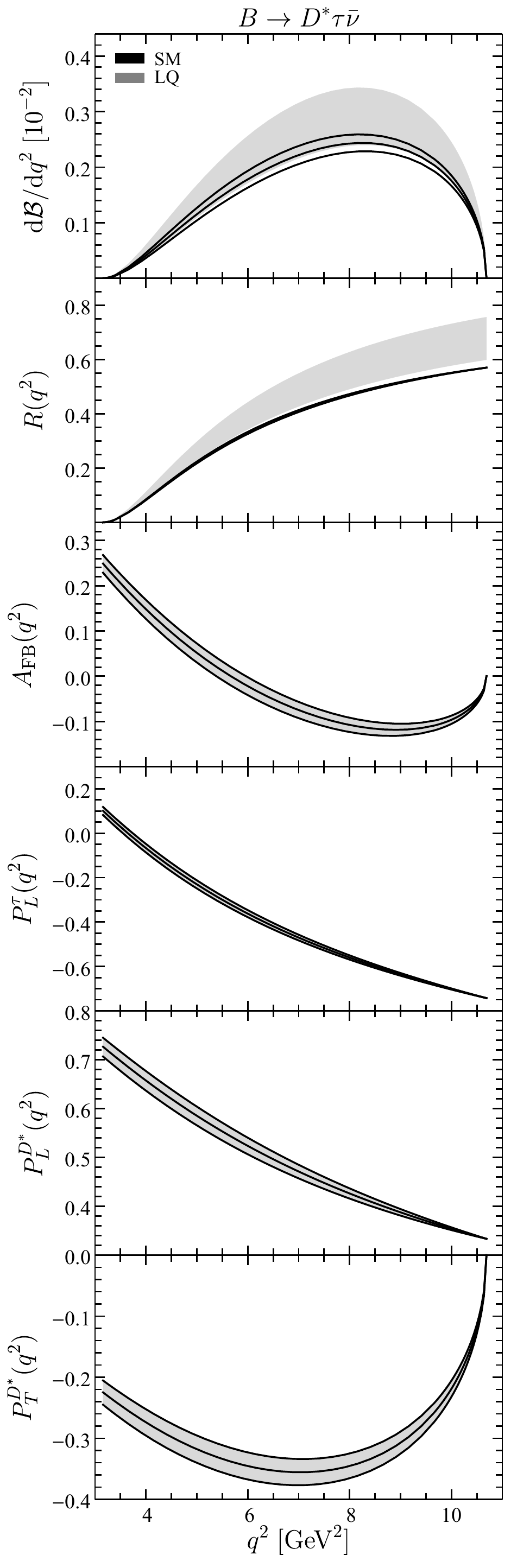}
    \qquad
    \includegraphics[width=0.39\linewidth]{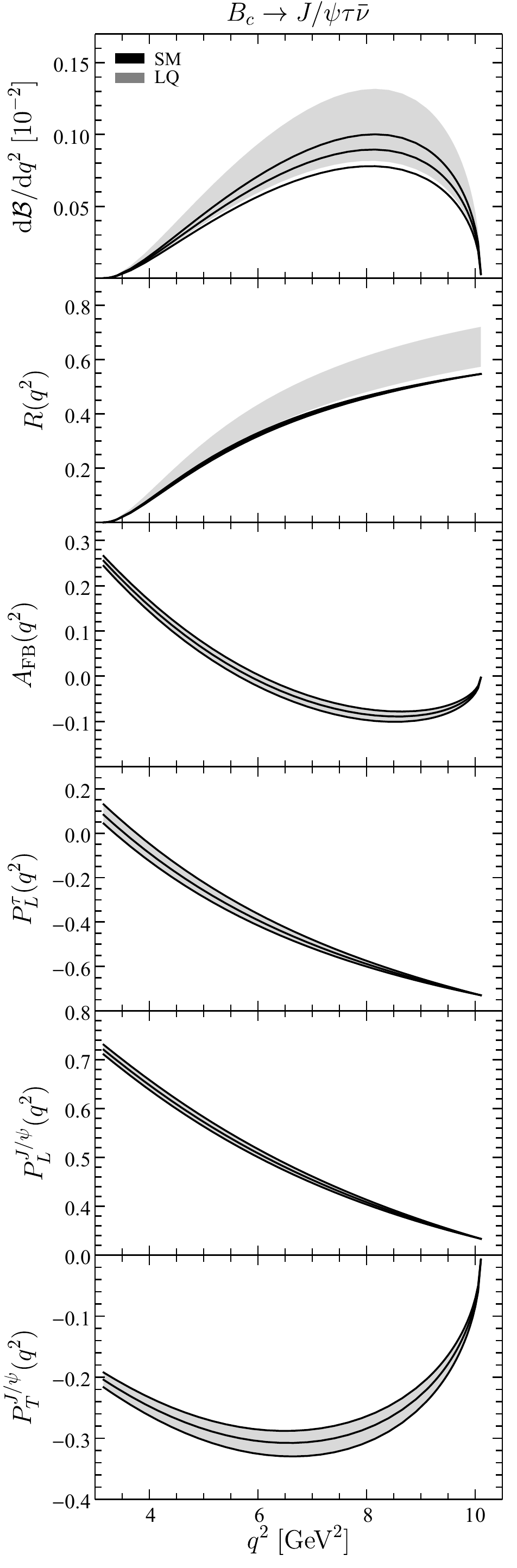}
    \caption{The $q^2$ distributions of the observables in $B \to D^* \tau \bar\nu$ (left) and $B_c \to J/\psi \tau \bar\nu$ (right) decays. Other captions are the same as in figure~\ref{fig:B2PS}.}
    \label{fig:B2V}
  \end{center}
\end{figure}

Now we start to analyze the $q^2$ distributions of the branching fraction $\mathcal B$, the LFU ratio $R$, the polarization fractions of the $\tau$ lepton ($P_L^\tau$) and the daughter hadron ($P_{L,T}^{D^{*}}$, $P_{L,T}^{J/\psi}$, $P_L^{\Lambda_c}$), as well as the lepton forward-backward asymmetry $A_{\rm FB}$. For the $B \to D \tau \nu$ and $B_c \to \eta_c \tau \bar\nu$ decays, both belonging to the ``$B\to P$'' transition, their differential observables in the SM and the LQ scenario are shown in figure~\ref{fig:B2PS}. It can be seen that all the differential observables of $B \to D \tau \bar\nu$ and $B_c \to \eta_c \tau \bar\nu$ decays are similar to each other, while the observables in the latter have larger theoretical uncertainties due to the less precise $B_c \to \eta_c$ transition form factors. Therefore, the $B \to D \tau \bar\nu$ decay is more sensitive to the LQ effects, with the differential branching fraction being largely enhanced, especially near $q^2 \sim 7\GeV^2$. The large difference between the SM and LQ predictions in this kinematic region could, therefore, provide a testable signature of the LQ effects. More interestingly, the $q^2$ distribution of the ratio $R$ in the LQ model is enhanced in the whole kinematic region and does not have overlap with the $1\sigma$ SM range. In the future, more precise measurements of these distributions are important to confirm the existence of possible NP effect in the $B \to D \tau \bar\nu$ decay. For the forward-backward asymmetry $\AFB$ and the $\tau$-lepton polarization fraction $P_L^\tau$ in both $B \to D \tau \bar\nu$ and $B_c \to \eta_c \tau\bar\nu$ decays, because the LQ effects only modify the Wilson coefficient $\mC_L^{\ell \nu_\ell}$, which is however canceled out exactly in the definitions of these observables (see eqs.~\eqref{eq:AFB} and \eqref{eq:PL}), the LQ predictions are indistinguishable from the SM ones, as shown in figure~\ref{fig:B2PS}. This feature is different from the NP scenarios that use scalar or tensor operators to explain the $R_{D^{(*)}}$ anomaly~\cite{Freytsis:2015qca,Bauer:2015knc,Li:2016pdv}.

The $q^2$ distributions of the observables in $B \to D^* \tau \bar\nu$ and $B_c \to J/\psi \tau \bar\nu$ decays are shown in figure~\ref{fig:B2V}. Since both of these two decays belong to ``$B \to V$'' transition, their differential observables are similar to each other. While the differential branching fractions of these two decays are enhanced in the LQ model, their theoretical uncertainties are larger than the ones in the $B \to D \tau \bar\nu$ decay. For the $q^2$ distributions of the ratios $R_{D^*}$ and $R_{J/\psi}$, they are largely enhanced in the whole kinematic region, especially in the large $q^2$ region. More importantly, although the ranges of the $q^2$-integrated ratio $R_{D^*,J/\psi}$ in the SM and the LQ scenario overlap at $1\sigma$ level, the $1\sigma$ ranges of the differential ratio $R_{D^*,J/\psi}(q^2)$ at large $q^2$ region in the SM and LQ show significant differences. The enhancements of $R_{D^*}$ and $R_{J/\psi}$ in the large $q^2$ region are stronger than the one observed in $R_D$. Measurements of the differential ratios in the large dilepton invariant mass region are, therefore, crucial to confirm the $R_{D^{(*)}}$ anomaly and to test the LQ model considered. Similarly to the ones in $B\to D \tau \bar\nu$ and $B_c \to \eta_c \tau \bar\nu$ decays, the angular distributions $\AFB$, $P_{L,T}^{D^*,J/\psi}$, and $P_L^\tau$ are also not affected by the LQ effects, as can be seen from figure~\ref{fig:B2V}.

\begin{figure}[t]
  \begin{center}
    \includegraphics[width=0.39\textwidth]{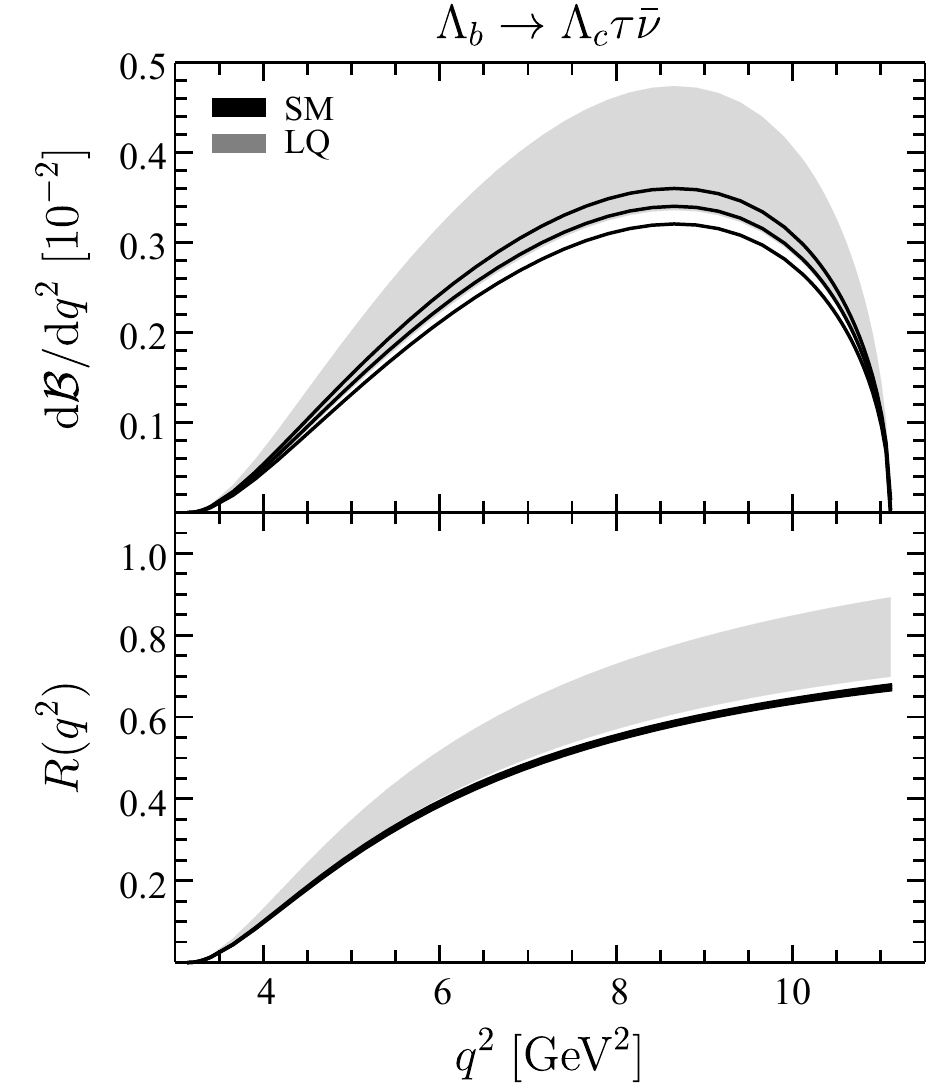}
    \qquad
    \includegraphics[width=0.39\textwidth]{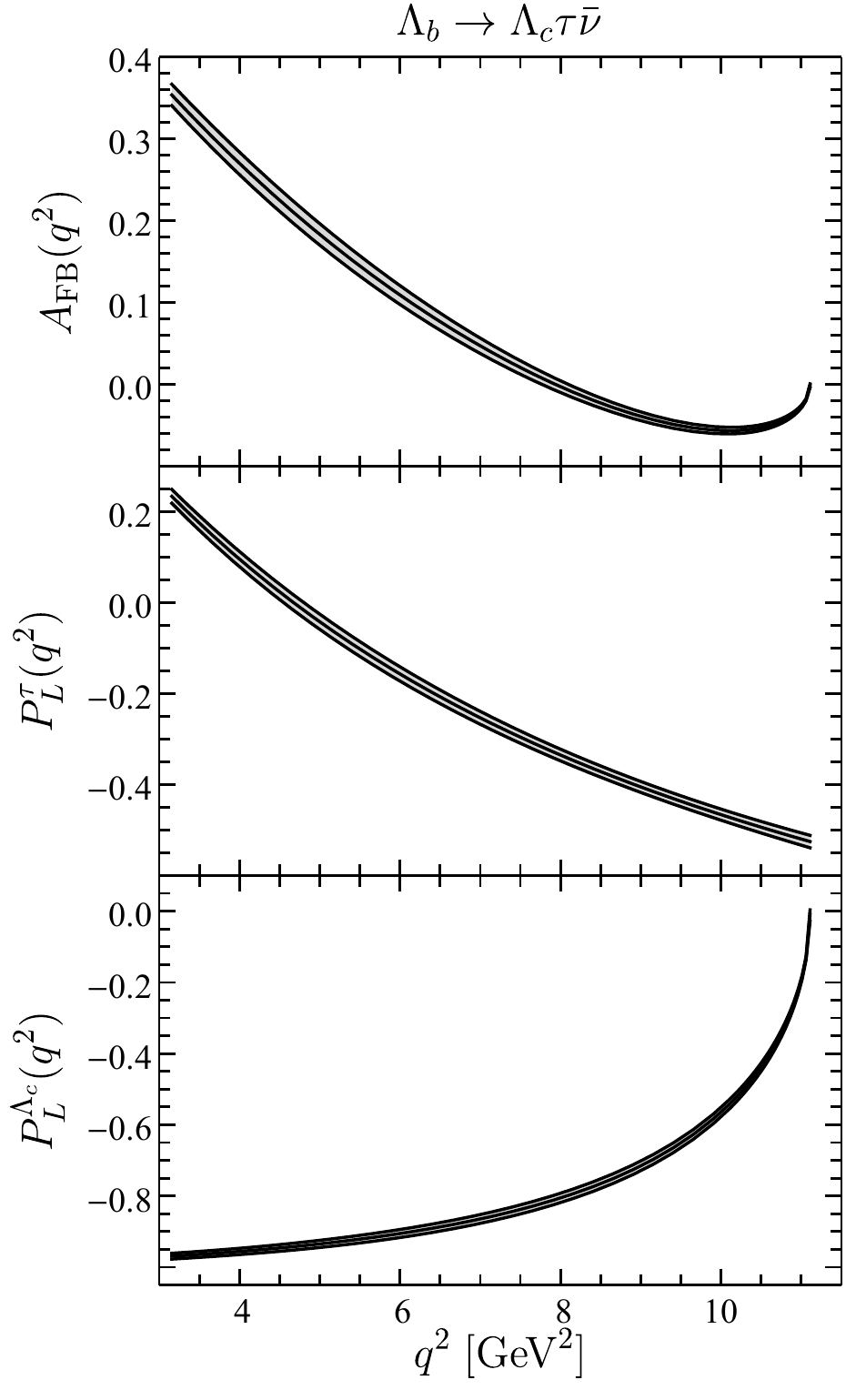}
    \caption{The $q^2$ distributions of the observables in $\Lambda_b \to \Lambda_c \tau \bar\nu$ decay. Other captions are the same as in figure~\ref{fig:B2PS}.}
    \label{fig:LamB2LamC}
  \end{center}
\end{figure}

For the $\Lambda_b \to \Lambda_c \tau \bar\nu$ decay, the $q^2$ distributions of the observables  are shown in figure~\ref{fig:LamB2LamC}. The situation is similar to the ones observed in $B \to D^* \tau \bar\nu$ and $B_c \to J/\psi \tau \bar\nu$ decays. The $q^2$ distributions of the branching fraction $\mathcal B$ and the ratio $R_{\Lambda_c}$ are largely enhanced by the LQ effects. At the large $q^2$ region, the differential ratio $R_{\Lambda_c}$ shows deviation between the $1\sigma$ allowed ranges of the SM and the LQ scenario. With large numbers of $\Lambda_b$ produced at the HL-LHC~\cite{Cerri:2018ypt}, we expect that this prediction could provide helpful information about the LQ effects. For the angular distributions, the LQ effects vanish due to the same reason as in the mesonic decays.

\section{Conclusions}
\label{sec:conclusions}

During the past few years, intriguing hints of LFU violation have emerged in the $B \to D^{(*)} \tau \bar\nu$ data. Motivated by the recent measurements of $R_{J/\psi}$, $P_L^\tau$, and $P_L^{D^*}$, we have revisited the LQ model proposed in ref.~\cite{Crivellin:2017zlb}, in which two scalar LQs, one being $SU(2)_L$ singlet and the other $SU(2)_L$ triplet, are introduced simultaneously. Taking into account the recent progresses on the transition form factors and the most up-to-date experimental data, we obtained constraints on the LQ couplings $\lambda_{23}^L$ and $\lambda_{33}^L$. Then, we investigated systematically the LQ effects on the five $b \to c \tau \bar\nu$ decays, $B \to D^{(*)}\tau \bar\nu$, $B_c \to \eta_c \tau\bar\nu$, $B_c \to J/\psi \tau \bar\nu$, and $\Lambda_b \to \Lambda_c \tau\bar\nu$. In particular, we have focused on the $q^2$ distributions of the branching fractions, the LFU ratios, and the various angular observables. Main results of this paper can be summarized as follows:
\begin{itemize}
\item After considering the $R_D$ and $R_{D^*}$ data, we obtain the bound on the LQ couplings, $0.03<\lambda_{23}^{L^*}\lambda_{33}^L<0.20$, at the $2\sigma$ level. It is found that the current measurements of $R_{J/\psi}$, $P_L^\tau$ and $P_L^{D^*}$ cannot provide further constraints on the LQ couplings.

\item The $B_s \to \tau^+ \tau^-$ decay is strongly correlated with $B \to D^{(*)} \tau \bar\nu$. In order to reproduce the $2\sigma$ experimental range of $R_{D^{(*)}}$, the LQ effects enhance $\mB(B_s \to \tau^+ \tau^-)$ by about $2$-$3$ orders of magnitude compared to the SM prediction, and hence reaches the expected sensitivity of the LHCb Upgrade II.
  
\item The differential branching fractions and the LFU ratios are largely enhanced by the LQ effects. Due to their small theoretical uncertainties, the latter provide testable signatures of the LQ model considered, especially in the large dilepton invariant mass squared region. It is also noted that $R_{\Lambda_c}$ in the baryonic decay $\Lambda_b \to \Lambda_c \tau \bar\nu$ has the potential to shed new light on the $R_{D^{(*)}}$ anomalies.
  
\item Since no new operators are generated by the LQ effects, all the angular distributions in the LQ model are the same as in the SM. We provide the most up-to-date SM predictions for the $\tau$-lepton forward-backward asymmetry, the $\tau$ and meson polarization fractions of the five $b \to c \tau \bar\nu$ modes. Although precision measurements of these angular distributions are very challenging at the HL-LHC and SuperKEKB, they are crucial to verify the LQ scenario investigated in this work.
\end{itemize}

The $q^2$ distributions of the branching fractions, the LFU ratios, and the various angular observables in $b \to c \tau \bar\nu$ transitions can help to confirm possible NP resolutions of the $R_{D^{(*)}}$ anomalies and to distinguish among the various NP candidates. With the experimental progresses expected from the SuperKEKB~\cite{Kou:2018nap} and the future HL-LHC~\cite{Cerri:2018ypt}, our predictions for these observables can be further probed in the near future.

\section*{Acknowledgements}
We thank Xin-Qiang Li for useful discussions. This work is supported by the National Natural Science Foundation of China under Grant Nos. 11775092, 11521064, 11435003, and 11805077. XY is also supported in part by the startup research funding from CCNU.
\paragraph{Note Added.}
After the completion of this work, the Belle Collaboration announced their  results of $R_{D}$ and $R_{D^*}$ with a semileptonic tagging method~\cite{Belle:Moriond,Abdesselam:2019dgh}. The measured values are $R_D^{\rm exp}=0.307\pm 0.037\,({\rm stat.})\pm 0.016\,({\rm syst.})$ and $R_{D^*}^{\rm exp}=0.283\pm 0.018 \, ({\rm stat.})\pm 0.014 \,({\rm syst.})$. After including this new measurement, the world averages become to $R_D^{\rm avg,\,2019}=0.337\pm 0.030$ and $R_{D^*}^{\rm avg,\,2019}=0.299\pm 0.013$~\cite{Murgui:2019czp}. The deviation of the current world averages from the SM predictions descreases from $3.8\sigma$ to $3.1\sigma$~\cite{Belle:Moriond}. Since the difference between the new and privous averages is small, our numerical results are expected to be qualitatively unchanged. For example, the updated bounds on $\lambda_{23}^{L*}\lambda_{33}^L$ in eq.~(\ref{eq:lam2333}) becomes to $-2.88<\lambda _{23}^{L*}\lambda _{33}^L <-2.73$ and $0.02<\lambda _{23}^{L*}\lambda _{33}^L<0.17$.
\begin{appendix}

	\section{Helicity amplitudes in $\boldsymbol{b \to c \tau \bar\nu}$ decays}
	\label{sec:helicity amplitude}
	
	In the presence of NP, the most general effective Hamiltonian for $b\to c\tau\bar{\nu}$ transition can be written as~\cite{Datta:2017aue,Sakaki:2014sea}
	\begin{align}
	{\cal{H}}_{\rm eff} =&  2\sqrt{2}G_F V_{cb}\Bigl[
	\bigl(1 + g_L\bigr) \bigl(\bar{c} \gamma_\mu P_L  b\bigr) \bigl(\bar\tau \gamma^\mu P_L \nu_\tau\bigr)
	+g_R  \bigl(\bar{c} \gamma_\mu P_R b\bigr) \bigl(\bar\tau \gamma^\mu P_L \nu_\tau\bigr)
	\nn\\
	& \qquad + \frac{1}{2}g_S \bigl(\bar{c}  b\bigr)  \bigl(\bar\tau P_L\nu_\tau\bigr)
	+ \frac{1}{2} g_P \bigl(\bar{c} \gamma_5 b \bigr) \bigl(\bar\tau P_L\nu_\tau\bigr)
	+ g_T \bigl(\bar{c}\sigma^{\mu \nu}P_Lb\bigr) \bigl(\bar{\tau}\sigma_{\mu \nu}P_L\nu_{\tau}\bigr)  \Bigr] + \text{h.c.}\,.
	\end{align}
	In this appendix, for completeness, we consider the most general case of NP and give the helicity amplitudes in the five $b \to c \tau \bar\nu$ decays, $B \to D^{(*)}\tau \bar\nu$, $B_c \to \eta_c \tau\bar\nu$, $B_c \to J/\psi \tau \bar\nu$, and $\Lambda_b \to \Lambda_c \tau\bar\nu$. Explicit expressions of the spinors and polarization vectors used to calculate the helicity amplitudes are also presented.

	\subsection{Kinematic conventions}
	
	To calculate the hadronic helicity amplitudes of $M\to N\tau \bar\nu$ in eq.~(\ref{eq:amp}), we work in the $M$ rest frame and follow the notation of ref.~\cite{Hagiwara:1989cu}:
	\begin{align}
	p^\mu_{M} =(m_{M},0,0,0),
	\qquad
	p^\mu_{N} =(E_{N},0,0,|\vec{p}_{N}|),
	\qquad
	q^{\mu}=(q_{0},0,0,-|\vec{q}\,|),
	\end{align}
	where $q^{\mu}$ is the four-momentum of the virtual vector boson in the $M$ rest frame, and
	\begin{align}
	q_0 =&\frac{1}{2m_{M}}(m_{M}^2-m_{N}^2+q^2),
	&E_{N}=&\frac{1}{2m_{M}}(m_{M}^2+m_{N}^2-q^2),\nn
	\\
	|\vec{q}\,| =&|\vec{p}_{N}|=\frac{1}{2m_{M}}\sqrt{Q_+Q_-},
	&Q_\pm =& (m_{M} \pm m_{N})^2 - q^2.
	\end{align}
	Then substituting the momentum into eq.~(\ref{eq:spinorU}), the Dirac spinors in the $\Lambda_b\to\Lambda_c\tau\nu_{\tau}$ decay can be written as
	\begin{align}
	u_{\Lambda_b}(\vec{p}_{\Lambda_b},\lambda_{\Lambda_b})=\sqrt{2m_{\Lambda_b}}
	\begin{spmatrix}
	\chi(\vec{p}_{\Lambda_b},\lambda_{\Lambda_b})
	\\
	0
	\end{spmatrix},
	\,\,
	u_{\Lambda_c}(\vec{p}_{\Lambda_c},\lambda_{\Lambda_c})=
	\begin{spmatrix}
	\sqrt{E+m_{\Lambda_c}}\chi(\vec{p}_{\Lambda_c},\lambda_{\Lambda_c})
	\\
	2\lambda_{\Lambda_c}\sqrt{E-m_{\Lambda_c}}\chi(\vec{p}_{\Lambda_c},\lambda_{\Lambda_c})
	\end{spmatrix},
	\end{align}
	where  $\chi(\vec{p}_{\Lambda_b},1/2)=\chi(\vec{p}_{\Lambda_c},1/2)=(1,0)^T,\chi(\vec{p}_{\Lambda_b},-1/2)=\chi(\vec{p}_{\Lambda_c},-1/2)=(0,1)^T$. 
	
	In the $B \to D^* \tau \bar\nu$ decay, the polarization vectors of the $D^*$ meson are given by
	\begin{align}
	\epsilon^{\mu}(\vec{p}_{D^*},0)=\frac{1}{m_{D^*}}\left(|\vec{p}_{D^*}|,0,0,E_{D^*}\right),
	\qquad
	\epsilon^{\mu}(\vec{p}_{D^*},\pm)=\frac{1}{\sqrt{2}}\left(0,\pm 1,i,0\right).
	\end{align}
	In all the five $b \to c \tau \bar\nu$ decays, the polarization vectors for the virtual vector boson $W$ can be written as
	\begin{align}\label{polvec}
	\epsilon^{\mu}(t) = \frac{1}{\sqrt{q^2}}\left(q_0,0,0,-|\vec{q}\,|\right),
	\quad
	\epsilon^{\mu}(0) = \frac{1}{\sqrt{q^2}}\left(|\vec{q}\,|,0,0,-q_0\right) ,
	\quad
	\epsilon^{\mu}(\pm) = \frac{1}{\sqrt{2}}\left(0,\mp 1,i,0\right),
	\end{align}
	and the orthonormality and completeness relation~\cite{Korner:1989qb}
	\begin{align}\label{eq:orthonormality and completeness}
	\sum\limits_\mu \epsilon_{\mu}^{*}(m)\epsilon^{\mu}(n)=g_{mn},
	\qquad
	\sum\limits_{m,n} \epsilon_{\mu}(m)\epsilon_{\nu}^{*}(n)g_{mn}=g_{\mu\nu},
	\qquad
	m,n\in\{t,\pm,0\},
	\end{align}
	where $g_{mn}=\text{diag}(+1,-1,-1,-1)$.

	In the calculation of the leptonic helicity amplitudes, we work in the rest frame of the virtual vector boson $W$, which is equivalent to the rest frame of the $\tau$-$\bar{\nu}_\tau$ system. Following ref.~\cite{Hagiwara:1989cu}, we have
	\begin{align}
	q^{\mu}=(\sqrt{q^2},0,0,0),
	\;\;
	p^\mu_{\tau} =(E_\tau ,|\vec{p}_\tau|\sin\theta_\tau,0,|\vec{p}_\tau|\cos\theta_\tau) ,
	\;\;
	p^\mu_{\bar\nu} = |\vec{p}_\tau | (1 ,-\sin\theta_\tau,0,-\cos\theta_\tau),
	\end{align}
	where $|\vec{p}_\tau| = \sqrt{q^2} v^2/2$, $E_\tau = |\vec{p}_\tau| + m_\tau^2/\sqrt{q^2}$, $v=\sqrt{1-m_\tau^2/q^2}$, and $\theta_\tau$ denotes the angle between the three-momenta of the $\tau$ and the $N$.  
	
	The Dirac spinors for $\tau$ and $\bar\nu_\tau$ read
	\begin{align}\label{eq:Dirac spinor}
	u_{\tau}(\vec{p}_\tau,\lambda_{\tau}) =
	\begin{spmatrix}
	\hphantom{2\lambda_\tau}\sqrt{E_\tau + m_\tau}\chi(\vec{p}_\tau,\lambda_{\tau})
	\\
	2\lambda_\tau\sqrt{E_\tau - m_\tau}\chi(\vec{p}_\tau,\lambda_{\tau})
	\end{spmatrix},
	\qquad
	v_{\bar{\nu}_\tau}(-\vec{p}_\tau,\frac{1}{2}) = \sqrt{E_\nu}
	\begin{spmatrix}
	\hphantom{-} \xi(-\vec{p}_\tau,\frac{1}{2})
	\\
	-\xi(-\vec{p}_\tau,\frac{1}{2})
	\end{spmatrix},
	\end{align}
	respectively. More details can be found in appendix~\ref{sec:Dirac spinor}
	%It is noted that the spinor $v_{{\bar\nu}_\tau}$ is obtained from the charge conjugate $v(\vec{p},s)\equiv C\bar{u}(\vec{p},s)^{T}$ with $C=i \gamma^0 \gamma^2$.

	The polarization vectors of the virtual vector boson in the $W$ rest frame are written as
	\begin{align}\label{polvec2}
	\bar{\epsilon}^{\mu}(t) = (1,0,0,0),
	\qquad
	\bar{\epsilon}^{\mu}(0) = (0,0,0,-1),
	\qquad
	\bar{\epsilon}^{\mu}(\pm) = \frac{1}{\sqrt{2}}(0,\mp 1,i,0),
	\end{align}
	which can also be obtained from eq.~(\ref{polvec}) by a Lorentz transformation and satisfy the orthonormality and completeness relation in eq.~(\ref{eq:orthonormality and completeness}).

	%In the above expressions, the two-component spinors follow the notation in ref.~\cite{Haber:1994pe} and have the following forms
	%\begin{align}
	%  \chi(\vec{p},+\frac{1}{2})=&
	%  \begin{spmatrix}
	%    \cos\frac{\theta}{2}
	%    \\
	%    e^{i\phi}\sin\frac{\theta}{2}
	%  \end{spmatrix},
	%  \qquad
	%  \chi(\vec{p},-\frac{1}{2})=
	%  \begin{spmatrix}
	%    -e^{-i\phi}\sin\frac{\theta}{2}
	%    \\
	%    \cos\frac{\theta}{2}
	%  \end{spmatrix},\nn\\
	%  \chi(-\vec{p},+\frac{1}{2})=&\left(\begin{array}{c} \sin(\frac{\theta}{2}) \\ -e^{i\phi}\cos(\frac{\theta}{2}) \end{array}\right),
	%  \qquad
	%  \chi(-\vec{p},-\frac{1}{2})=\left(\begin{array}{c} e^{-i\phi}\cos(\frac{\theta}{2}) \\\sin(\frac{\theta}{2}) \end{array}\right)
	%\end{align}
	%for $\vec{p}=(\sin\theta\cos\phi,\sin\theta\sin\phi,\cos\theta)$. And $\xi(\pm\vec{p}, s)$ can be obtained from the relation $\xi(\vec{p}, \pm 1/2)= \pm \chi(\vec{p}, \mp 1/2)$ and $\xi(-\vec{p}, \pm 1/2)= \pm \chi(-\vec{p}, \mp 1/2)$. 
	
	\subsection{Dirac spinor}
	\label{sec:Dirac spinor}
	
	The definitions of the helicity operator $h_{\vec{p}}$ and its eigenstates are given as follows~\cite{Haber:1994pe}
	\begin{align*}
	h_{\vec{p}}\equiv\frac{1}{2}\hat{\vec{p}}\cdot\vec{\sigma},\quad\hat{\vec{p}}\equiv\frac{\vec{p}}{|\vec{p}|},\quad h_{\vec{p}}~\chi(\vec{p},s)=s~\chi(\vec{p},s),
	\end{align*}
	where $\vec{p}$ denotes the momentum of the particle and $\vec{\sigma}=\{\sigma^1,\sigma^2,\sigma^3\}$ the Pauli matrices. Eigenstates of the helicity operator $h_{\vec{p}}$ read 
	\begin{align}
	\chi(\vec{p},\frac{1}{2})=&
	\begin{spmatrix}
	\cos\frac{\theta}{2} \\ e^{i\phi}\sin\frac{\theta}{2}
	\end{spmatrix},
	&\chi(\vec{p},-\frac{1}{2})=&
	\begin{spmatrix}
	-e^{-i\phi}\sin\frac{\theta}{2} \\ \cos\frac{\theta}{2}
	\end{spmatrix},\nn\\
	\chi(-\vec{p},\frac{1}{2})=&
	\begin{spmatrix}
	\sin\frac{\theta}{2} \\ -e^{i\phi}\cos\frac{\theta}{2}
	\end{spmatrix},
	&\chi(-\vec{p},-\frac{1}{2})=&
	\begin{spmatrix}
	e^{-i\phi}\cos\frac{\theta}{2} \\\sin\frac{\theta}{2}
	\end{spmatrix},
	\end{align}
	for the normalized momentum $\hat{\vec{p}}=\{\sin\theta\cos\phi,\sin\theta\sin\phi,\cos\theta\}$.  
	
	Using these eigenstates, solution of Dirac equation $(\gamma^{\mu} p_{\mu}-m)u(\vec{p},s)=0$ in Dirac representation can be written as
	\begin{align}\label{eq:spinorU}
	u(\vec{p},s)=
	\begin{spmatrix}
	\hphantom{2s}\sqrt{E+m}~\chi(\vec{p},s)
	\\
	2s\sqrt{E-m}~\chi(\vec{p},s)
	\end{spmatrix}.
	\end{align} 
	Then, spinor for antiparticle can be obtained by $v(\vec{p},s)\equiv C\bar{u}(\vec{p},s)^{T}=i\gamma^0\gamma^2 \bar{u}(\vec{p},s)^{T}$\footnote{The selection  $C=i\gamma^2\gamma^0$ is also permissible, but the $v(\vec{p},s)$ will have an additional negative sign. }, whose explicit expression reads
	\begin{align}
	\label{eq:spinorV}
	v(\vec{p},s)=
	\begin{spmatrix}
	\hphantom{-}\sqrt{E-m}~\xi(\vec{p},s)\\ -2s\sqrt{E+m} ~\xi(\vec{p},s)
	\end{spmatrix},
	\end{align}
	where $\xi(\vec{p},s)=\chi(\vec{p},-s)$ and $\xi(\vec{p},s)$ satisfies $h_{\vec{p}}~\xi(\vec{p},s)=-s~\xi(\vec{p},s)$. 
	
	The spinors in Weyl representation read
	\begin{align}
	u_W(\vec{p},s)=
	\begin{spmatrix}
	\sqrt{E-2s\abs{\vec{p}}}~\chi(\vec{p},s)\\ \sqrt{E+2s\abs{\vec{p}}} ~\chi(\vec{p},s)
	\end{spmatrix},
	\quad
	v_W(\vec{p},s)= 
	\begin{spmatrix}
	-2s\sqrt{E+2s\abs{\vec{p}}}~\xi(\vec{p},s)\\ \hphantom{-} 2s\sqrt{E-2s\abs{\vec{p}}} ~\xi(\vec{p},s)
	\end{spmatrix}.
	\end{align}
	They can also be obtained from Dirac representation by the relation $u_{W}(\vec{p},s)=X u(\vec{p},s)$ with the transformation matrix
	\begin{align*}
	X=\frac{1}{\sqrt{2}}
	\begin{spmatrix}
	1&-1\\1&1
	\end{spmatrix}.
	\end{align*}
	In the $\tau$-$\bar\nu_\tau$ center-of-mass frame, we emphasize that if the $\tau$ spinor is specified as $u(\vec{p},s)$ in leptonic helicity amplitude,  then the $\bar\nu_\tau$ spinor has the form $v(-\vec{p},s)$, as in eq.~(\ref{eq:Dirac spinor}). All calculations in our work are in Dirac representation. 
	
	\subsection{Leptonic helicity amplitudes}
	\label{sec:Leptonic  helicity amplitudes}
	
	The leptonic helicity amplitudes in eq.~(\ref{eq:amp}) are defined as~\cite{Hagiwara:1989cu}
	\begin{align}
	L^{SP}_{\lambda_\tau}=&\bra{\tau\bar{\nu}_\tau}\bar{\tau} (1-\gamma_5)\nu_\tau\ket{0}=\bar{u}_\tau(\vec{p}_\tau,\lambda_{\tau})(1-\gamma_5)v_{\bar{\nu}_\tau}(-\vec{p}_\tau,1/2),
	\nn\\
	L^{VA}_{\lambda_\tau,\la_W}=&\bar{\epsilon}^{\mu} (\la_W)\bra{\tau\bar{\nu}_\tau}\bar{\tau}\gamma_\mu (1-\gamma_5)\nu_\tau\ket{0} =\bar{\epsilon}^{\mu}(\la_W)\bar{u}_\tau(\vec{p}_\tau,\lambda_{\tau})\gamma_\mu (1-\gamma_5)v_{\bar{\nu}_\tau}(-\vec{p}_\tau, 1/2) , \nn\\
	L^{T}_{\lambda_\tau,\la_{W_1} ,\la_{W_2}} =&-i\bar{\epsilon}^{\mu} (\la_{W_1})\bar{\epsilon}^{\nu} (\la_{W_2})\bra{\tau\bar{\nu}_\tau}\bar{\tau}\sigma_{\mu \nu} (1-\gamma_5)\nu_\tau\ket{0} \nn\\
	=&-i\bar{\epsilon}^{\mu} (\la_{W_1})\bar{\epsilon}^{\nu} (\la_{W_2})\bar{u}_\tau(\vec{p}_\tau,\lambda_{\tau})\sigma_{\mu \nu} (1-\gamma_5)v_{\bar{\nu}_\tau}(-\vec{p}_\tau, 1/2),
	\end{align}
	It is straightforward to obtain $L^T_{\lambda_\tau,\la_{W_1} ,\la_{W_2}}=-L^T_{\lambda_\tau ,\la_{W_2},\la_{W_1}}$. The non-zero leptonic helicity amplitudes read
	\begin{align}
	L^{SP}_{1/2}&= 2\sqrt{q^2}v,&L^{VA}_{1/2,t}&=2m_\tau v,\nn\\
	L^{VA}_{1/2,0}&=-2m_\tau v \cos\theta_\tau ,&L^{VA}_{-1/2,0}&=2\sqrt{q^2}v \sin\theta_\tau,\nn\\
	L^{VA}_{1/2,\pm}&=\mp\sqrt{2} m_\tau v \sin\theta_\tau,&L^{VA}_{-1/2,\pm}&=\sqrt{2 q^2}v (-1 \mp \cos\theta_\tau),\nn\\
	L^{T}_{1/2,0,\pm} &= \pm L^{T}_{1/2,\pm,t} = \sqrt{2 q^2}v\sin\theta_\tau,
	&L^{T}_{1/2,t,0}&=L^{T}_{1/2,+,-}=-2\sqrt{q^2}v \cos\theta_\tau,\nn\\
	L^{T}_{-1/2,0,\pm}&= \pm L^{T}_{-1/2,\pm,t}  = \sqrt{2}m_\tau v (\pm 1+ \cos\theta_\tau),
	&L^{T}_{-1/2,t,0}&=L^{T}_{-1/2,+,-}=2m_\tau v \sin\theta_\tau.
	\end{align}
	\subsection{Hadronic helicity amplitudes}
	
	The hadronic helicity amplitudes $M \to N$ are defined as
	\begin{align}
	H^S_{\la_M,\la_N}&=\bra{N(\la_N)}\bar{c} b\ket{M(\la_M)},\nn\\
	H^P_{\la_M,\la_N}&=\bra{N(\la_N)}\bar{c}\gamma_5 b\ket{M(\la_M)},\nn\\
	H^V_{\la_M,\la_N,\la_{W}}&=\,\epsilon_\mu^{*}(\la_{W})\bra{N(\la_N)}\bar{c}\gamma^\mu b\ket{M(\la_M)}, \nn\\
	H^A_{\la_M,\la_N,\la_{W}}&=\,\epsilon_\mu^{*}(\la_{W})\bra{N(\la_N)}\bar{c}\gamma^\mu\gamma_5 b\ket{M(\la_M)},\nn\\
	H^{T_1,\la_M}_{\la_N,\la_{W_1} ,\la_{W_2}}&=i\epsilon_\mu^{*}(\la_{W_1})\epsilon_\nu^{*}(\la_{W_2})\bra{N(\la_N)}\bar{c}\sigma^{\mu \nu} b\ket{M(\la_M)}, \nn\\
	H^{T_2,\la_M}_{\la_N,\la_{W_1} ,\la_{W_2}}&=i\epsilon_\mu^{*}(\la_{W_1})\epsilon_\nu^{*}(\la_{W_2})\bra{N(\la_N)}\bar{c}\sigma_{\mu \nu}\gamma_5 b\ket{M(\la_M)},
	\end{align}
	and
	\begin{align}\label{eq:helicity amplitude}
	H^{SP}_{\la_M,\la_N}&=g_S H^S_{\la_M,\la_N}+g_P H^P_{\la_M,\la_N}, \nn\\
	H^{VA}_{\la_M,\la_N,\la_{W}}&=(1+g_L+g_R)H^V_{\la_N,\la_{W}}-(1+g_L-g_R)H^A_{\la_M,\la_N,\la_{W}}, \nn\\
	H^{T,\la_M}_{\la_N,\la_{W_1} ,\la_{W_2}}&=g_TH^{T_1,\la_M}_{\la_N,\la_{W_1} ,\la_{W_2}}-g_T H^{T_2,\la_M}_{\la_N,\la_{W_1} ,\la_{W_2}} ,
	\end{align}
	It is straightforward to obtain $H^{T,\la_M}_{\la_N,\la_{W_1} ,\la_{W_2}}=-H^{T,\la_M}_{\la_N,\la_{W_2} ,\la_{W_1}}$. The amplitudes $H_{\lambda_N,\lambda_{W_1},\lambda_{W_2}}^{T_1,\lambda_M}$ and $H_{\lambda_N,\lambda_{W_1},\lambda_{W_2}}^{T_2,\lambda_M}$ are connected by the relation $\sigma_{\mu \nu}\gamma_{5} =-(i/2)\epsilon^{\mu\nu\alpha\beta}\sigma_{\alpha\beta}$, where $\epsilon^{0123}=-1$.

	\section{Form factors}
	\label{sec:form factor}
	
	The hadronic matrix elements for $B\to D$ transition can be parameterized in terms of form factors $F_{+,0,T}$~\cite{Sakaki:2013bfa,Bardhan:2016uhr}. In the BGL parametrization, the form factors $F_{+,0}$ can be written as expressions of $a_n^+$ and $a_n^0$~\cite{Bigi:2016mdz},
	\begin{align}
	F_{+}(z) =\frac{1}{P_{+}(z)\phi_{+}(z,\mathcal{N})}\sum_{n=0}^{\infty} a_{n}^{+}z^n(w,\mathcal{N}),
	\qquad
	F_{0}(z) =\frac{1}{P_{0}(z)\phi_{0}(z,\mathcal{N})}\sum_{n=0}^{\infty} a_{n}^{0}z^n(w,\mathcal{N}),
	\end{align}
	where $r=m_D / m_B$, $\mathcal{N}=(1+r)/(2\sqrt{r})$, $w=(m_{B}^2+m_{D}^2-q^{2})/(2m_{B}m_{D})$, $z(w,\mathcal{N})=(\sqrt{1+w}-\sqrt{2\mathcal{N}})/(\sqrt{1+w}+\sqrt{2\mathcal{N}})$, and $F_+(0)=F_0(0)$. The values of the fit parameters    are taken from ref.~\cite{Bigi:2016mdz}. Expressions of the tensor form factor $F_T$ can be found in ref.~\cite{Sakaki:2013bfa}.

	For $B\to D^*$ transition, the relevant form factors $\{V,A_{0,1,2}\}$ can be written in terms of the form factors $\{h_V,h_{A_{1,2,3}}\}$ in the Heavy Quark Effective Theory (HQET)~\cite{Sakaki:2013bfa}, 
	\begin{align}
	V(q^2) &= { m_+ \over 2\sqrt{m_B m_{D^{*}}} } \, h_V(w),  \nonumber\\
	A_0(q^2) &= { 1 \over 2\sqrt{m_B m_{D^{*}}} } \left[ { m_+^2 - q^2 \over 2m_{D^{*}} } \, h_{A_1}(w) - { m_+m_- + q^2 \over 2m_B } \, h_{A_2}(w) - {m_+m_-- q^2 \over 2m_{D^{*}} } \, h_{A_3}(w) \right] ,\nonumber
	\\
	A_1(q^2) &= { m_+^2 - q^2 \over 2\sqrt{m_B m_{D^{*}}} m_+  } \, h_{A_1}(w), \nonumber
	\\
	A_2(q^2) &= {m_+ \over 2\sqrt{m_B m_{D^{*}}} } \left[ h_{A_3}(w) + { m_{D^{*}} \over m_B } h_{A_2}(w) \right] ,
	\end{align}
	where $m_\pm = m_B \pm m_{D^*}$ and $w=(m_B^2+m_{D^*}^2-q^2)/2m_Bm_{D^*}$. In the CLN parametrization, the HQET form factors can be expressed as~\cite{Caprini:1997mu}
	\begin{align}
	\frac{h_V(w)}{h_{A_1}(w) } = R_1(w) ,
	\qquad
	\frac{h_{A_2}(w)}{h_{A_1}(w)} = { R_2(w)-R_3(w) \over 2\,r_{D^{*}} },
	\qquad
	\frac{h_{A_3}(w)}{h_{A_1}(w)} &= { R_2(w)+R_3(w) \over 2 } ,   
	\end{align}
	with $r=m_{D^*}/m_B$. Numerically we have, 
	\begin{align}
	h_{A_1}(w)&=h_{A_1}(1)[1-8\rho_{D^*}^2 z + (53 \rho_{D^*}^2 -15)z^2 - (231 \rho_{D^*}^2 -91)z^3],\nonumber\\
	R_1(w)&=R_1(1)-0.12(w-1)+0.05(w-1)^2,\nonumber\\
	R_2(w) &=R_2(1)+0.11(w-1)-0.06(w-1)^2,\nonumber\\
	R_3(w) &=1.22 -0.052(w-1) +0.026(w-1)^2,
	\end{align}
	with $z=(\sqrt{w+1}-\sqrt{2})/(\sqrt{w+1}+\sqrt{2})$.  The fit parameters $R_{1}(1)$, $R_{2}(1)$, $h_{A_1}(1)$ and $\rho_{D^{*}}^2$ are taken from ref.~\cite{Jaiswal:2017rve}. Expressions of the tensor form factors $T_{1,2,3}$ can be found in ref.~\cite{Sakaki:2013bfa}.

	The $\Lambda_b\rightarrow\Lambda_c$ hadronic matrix elements can be written in terms of ten helicity form factors $\{F_{0,+,\perp},G_{0,+,\perp},h_{+,\perp},\widetilde{h}_{+,\perp}\}$~\cite{Detmold:2015aaa,Datta:2017aue}. Following ref.~\cite{Detmold:2015aaa}, the lattice calculations are fitted to two (Bourrely-Caprini-Lellouch) BCL $z$-parametrization. In the so called ``nominal'' fit, a form factor $f$ reduces to the form
	\begin{align}\label{eq:nominalfitphys}
	f(q^2) = \frac{1}{1-q^2/(m_{\rm pole}^f)^2} \big[ a_0^f + a_1^f\:z^f(q^2)  \big], 
	\end{align}
	while a form factor $f$ in the higher-order fit is given by
	\begin{align}\label{eq:HOfitphys}
	f_{\rm HO}(q^2) =& \frac{1}{1-q^2/(m_{\rm pole}^f)^2} \bigl\lbrace a_{0,{\rm HO}}^f + a_{1,{\rm HO}}^f\:z^f(q^2) + a_{2,{\rm HO}}^f\:[z^f(q^2)]^2  \bigr\rbrace,
	\end{align}
	where $t_0 = (m_{\Lambda_b} - m_{\Lambda_c})^2$, $t_+^f = (m_{\rm pole}^f)^2$, and $  z^f(q^2) = (\sqrt{t_+^f-q^2}-\sqrt{t_+^f-t_0})/(\sqrt{t_+^f-q^2}+\sqrt{t_+^f-t_0})$.The values of the fit parameters and all the pole masses are taken from ref.~\cite{Datta:2017aue}.

	In addition, the form factors for $B_c \to J/\psi\ell\bar{\nu_\ell} $ and $B_c \to \eta_c\ell\bar{\nu_\ell}$ decays are taken form the results in the Covariant Light-Front Approach in ref.~\cite{Wang:2008xt}.

	\section{Observables in $\boldsymbol{b \to c \tau \bar\nu}$ decays}
	\label{sec:obs}

	\subsection{$\boldsymbol{B \to D \tau \bar\nu}$ and $\boldsymbol{B_c \to \eta_c \tau \bar\nu}$ decays}
	\label{sec:B2D}
	
	Since similar expressions hold for the $B \to D \tau \bar\nu$ and $B_c \to \eta_c \tau \bar\nu$ decays, we only give the theoretical formulae of the former. Using the form factors in appendix~\ref{sec:form factor}, the non-zero helicity amplitudes for the $B \to D \tau \bar\nu$ decay in eq.~(\ref{eq:helicity amplitude}) can be written as 
	\begin{align}
	H_{0}^{VA}(q^2) &=(1+g_L+g_R)\sqrt{\frac{Q_+Q_-}{q^2}} F_{+}(q^2) ,
	&
	H_{t}^{VA}(q^2) &=(1+g_L+g_R)\frac{m_B^2-m_D^2}{\sqrt{q^2}} F_0(q^2) ,\nonumber\\
	H^{SP}(q^2) &=g_S{m_B^2-m_D^2 \over m_b-m_c} F_0(q^2)  ,
	&
	H_{-,+}^T(q^2) &= H_{t,0}^T(q^2) =g_T{\sqrt{Q_+Q_-} \over m_B+m_D} F_T(q^2) .
	\end{align}
	Then, the differential decay width in eq.~(\ref{eq:dga}) and angular observables in eq.~(\ref{eq:AFB}) and (\ref{eq:PL}) are obtained
	\begin{align}
	\frac{{\rm d}\Gamma}{{\rm d}q^2}=&\frac{N_D}{2} \biggl[\frac{3m_{\tau}^2}{q^2}|H^{VA}_{t}|^2+ \Bigl(2+\frac{m_\tau^2}{q^2}\Bigr) |H^{VA}_{0}|^2+3|H^{SP}|^2 +16\Big(1+\frac{2m_\tau^2}{q^2}\Big)|H^{T}_{t,0}|^2
	\nn\\
	&\hspace{5em} +\frac{6m_\tau}{\sqrt{q^2}} \Re[H^{SP}H^{VA*}_{t}]+\frac{24m_\tau}{\sqrt{q^2}} \Re[H^{T}_{t,0}H^{VA*}_{0}] \biggr], \\[1em]
	\frac{{\rm d}A_{\rm FB}}{{\rm d} q^2}=&\frac{3N_D}{2}\Re \biggl[\biggl(4H^{T*}_{t,0}+ \frac{m_\tau}{\sqrt{q^2}}H_0^{VA*}\biggr) \biggl(H^{SP} + \frac{m_\tau}{\sqrt{q^2}}H_t^{VA} \biggr) \biggr],
	\\[1em]
	\frac{{\rm d}P_{L}^{\tau}}{{\rm d} q^2}=&\frac{1}{\rm{d}\Gamma/dq^2}\frac{N_D}{2} \biggl[\frac{3m_\tau^2}{q^2}|H^{VA}_{t}|^2+ \Bigl(\frac{m_\tau^2}{q^2}-2 \Bigr)|H^{VA}_{0}|^2  +3|H^{SP}|^2+16 \Bigl(1-\frac{2m_\tau^2}{q^2} \Bigr)|H^{T}_{t,0}|^2  
	\nn\\
	& \hspace{5em} +\frac{6 m_\tau}{\sqrt{q^2}} \Re[H^{SP}H^{VA*}_{t}]-\frac{8 m_\tau}{\sqrt{q^2}}\Re[H^{T}_{t,0}H^{VA*}_{0}] \biggr] ,
	\end{align}
	with
	\begin{align}
	N_D=\frac{G_{F}^{2}|V_{cb}|^2}{192\pi^{3}}\frac{q^2\sqrt{Q_+ Q_-}}{m_{B}^3}\Big(1-\frac{m_\tau^2}{q^2}\Big)^2.
	\end{align}

	\subsection{$\boldsymbol{B \to D^* \tau \bar\nu}$ and $\boldsymbol{B_c \to J/\psi \tau \bar\nu}$ decays}
	\label{sec:B2DV}
	
	Since similar expressions hold for the $B \to D^* \tau \bar\nu$ and $B_c \to J/\psi \tau \bar\nu$ decays, only theoretical formulae of the former are given in this subsection. Using the form factors in appendix~\ref{sec:form factor}, the non-zero helicity amplitudes for the $B \to D^* \tau \bar\nu$ decay in eq.~(\ref{eq:helicity amplitude}) can be written as
	\begin{align}
	H_0^{SP}(q^2) &=-g_P{ \sqrt{Q_+Q_-} \over m_b+m_c } A_0(q^2) ,
	\nn\\
	H^{VA}_{\pm,\pm}(q^2) &=-(1+g_L-g_R) m_+ A_1(q^2) \pm (1+g_L+g_R){ \sqrt{Q_+Q_-} \over m_+ } V(q^2) ,
	\nn\\
	H^{VA}_{0,t}(q^2) &=-(1+g_L-g_R){ \sqrt{Q_+Q_-} \over \sqrt{q^2} } A_0(q^2) , 
	\nn\\
	H^{VA}_{0,0}(q^2) &=\frac{(1+g_L-g_R)}{2m_\Dst\sqrt{q^2}}\biggl[- m_+(m_+ m_- -q^2) A_1(q^2)  +{Q_+Q_- \over m_+ } A_2(q^2)\biggr] ,
	\nn\\
	H^{T}_{\pm,\pm,t}(q^2) =&\pm H^{T}_{\pm,\pm,0}(q^2)=\frac{g_{T}}{\sqrt{q^{2}}} \biggl[\mp\sqrt{Q_+Q_-} T_1(q^2)-m_+m_- T_2(q^2) \biggr] ,
	\nn\\
	H^{T}_{0,t,0}(q^2) =&~H^{T}_{0,-,+}(q^2)=\frac{g_T}{2m_{D^{*}}}\biggl[-(m_B^2+3m_{D^{*}}^2-q^2) T_2(q^2)+\frac{Q_+Q_-}{m_+ m_-}T_3(q^2) \biggr] ,
	\end{align}
	with $m_\pm = m_B \pm m_{D^*}$. Then, the differential decay width in eq.~(\ref{eq:dga}) and the angular observables in eq.~(\ref{eq:AFB}) and (\ref{eq:PL}) are obtained, respectively, as
	\begin{align}
	\frac{{\rm d}\Gamma}{{\rm d} q^2}=&N_{D^*} \biggl[\frac{3m_\tau^2}{2q^2} |H^{VA}_{0,t}|^2+ \Bigl( 1+\frac{m_\tau^2}{2q^2} \Bigr)(|H^{VA}_{-,-}|^2+|H^{VA}_{0,0}|^2+|H^{VA}_{+,+}|^2)\nn 
	\\ 
	&+\frac{3}{2}|H^{SP}_{0}|^2 +8 \Bigl( 1+\frac{2 m_\tau^2}{q^2} \Bigr)(|H^{T}_{0,t,0}|^2+|H^{T}_{+,+,t}|^2+|H^{T}_{-,-,t}|^2) \nn
	\\
	&+ \frac{3m_\tau}{\sqrt{q^2}}\Re[H^{SP}_{0} H^{VA*}_{0,t}]+\frac{12 m_\tau}{\sqrt{q^2}}(\Re[H^{T}_{0,t,0}H^{VA*}_{0,0}-H^{T}_{+,+,t}H^{VA*}_{+,+}-H^{T}_{-,-,t}H^{VA*}_{-,-}])\biggr],
	\\[1em]
	\frac{{\rm d}A_{\rm FB}}{{\rm d} q^2}=&\frac{3N_{D^*}}{4} \biggl[\frac{2m_\tau^2}{q^2} \Re[H^{VA}_{0,0}H^{VA*}_{0,t}]-|H^{VA}_{-,-}|^2+|H^{VA}_{+,+}|^2+8\Re[H_0^{SP} H_{0,t,0}^{T*}] \nn\\
	&+\frac{16m_\tau^2}{q^2}(|H^{T}_{+,+,t}|^2-|H^{T}_{-,-,t}|^2)+\frac{2m_\tau}{\sqrt{q^2}}\Re[H^{SP}_{0}H^{VA*}_{0,0}]\nn
	\\
	&+\frac{8m_\tau}{\sqrt{q^2}}\Re[H^{T}_{0,t,0}H^{VA*}_{0,t}+H^{T}_{-,-,t}H^{VA*}_{-,-}-H^{T}_{+,+,t}H^{VA*}_{+,+}] \biggr],
	\\[1em]
	\frac{{\rm d}P_{L}^{\Dst}}{{\rm d} q^2} =&\frac{1}{{\rm d}\Gamma/{\rm d}q^2}\frac{N_{D^*}}{2}\biggl[\frac{3m_\tau^2}{q^2}|H^{VA}_{0,t}|^2+ \Bigl( 2 + \frac{m_\tau^2}{q^2} \Bigr) |H^{VA}_{0,0}|^2 +3|H^{SP}_{0}|^2 +16 \Bigl(1+\frac{2m_\tau^2}{q^2} \Bigr)|H^{T}_{0,t,0}|^2 \nn
	\\
	& +\frac{6 m_\tau}{\sqrt{q^2}}\Re[H^{SP}_{0}H^{VA*}_{0,t}]+\frac{24 m_\tau}{\sqrt{q^2}}\Re[H^{T}_{0,t,0}H^{VA*}_{0,0}] \biggr],
	\\[1em]
	\frac{{\rm d} P_L^\tau}{{\rm d}q^2}=&\frac{1}{{\rm d}\Gamma/{\rm d}q^2}\frac{N_{D^*}}{2}\biggl[\frac{3m_\tau^2}{q^2}|H^{VA}_{0,t}|^2+\Bigl(\frac{m_\tau^2}{q^2}-2\Bigr)(|H^{VA}_{+,+}|^2+|H^{VA}_{0,0}|^2+|H^{VA}_{-,-}|^2 )\nn
	\\
	&+3|H^{SP}_{0}|^2+\frac{6 m_\tau}{\sqrt{q^2}}\Re[H^{SP}_{0}H^{VA*}_{0,t}]+16 \Bigl(1-\frac{2m_\tau^2}{q^2} \Bigr)(|H^{T}_{0,t,0}|^2+|H^{T}_{-,-,t}|^2+|H^{T}_{+,+,t}|^2) \nn
	\\
	& +\frac{8m_\tau}{\sqrt{q^2}}\Re[H^{T}_{-,-,t}H^{VA*}_{-,-}+H^{T}_{+,+,t}H^{VA*}_{+,+}-H^{T}_{0,t,0}H^{VA*}_{0,0}] \biggr],
	\\[1em]
	\frac{{\rm d} P_{T}^{\Dst}}{ {\rm d} q^2}=&\frac{1}{{\rm d}\Gamma/{\rm d} q^2}\frac{N_{D^*}}{2}\biggl[\Bigl(2 + \frac{m_\tau^2}{q^2} \Bigr)(|H^{VA}_{+,+}|^2-|H^{VA}_{-,-}|^2) +16(1+\frac{2m_\tau^2}{q^2})(|H^{T}_{+,+,t}|^2-|H^{T}_{-,-,t}|^2)\nn\\ 
	&+\frac{24m_\tau}{\sqrt{q^2}}\Re[H^{T}_{-,-,t}H^{VA*}_{-,-}-H^{T}_{+,+,t}H^{VA*}_{+,+}] \biggr],
	\end{align}
	with
	\begin{align}
	N_{D^*}=\frac{G_{F}^{2}|V_{cb}|^2}{192\pi^{3}}\frac{q^2\sqrt{Q_+ Q_-}}{m_{B}^3}\Big(1-\frac{m_\tau^2}{q^2}\Big)^2.
	\end{align}

	\subsection{$\boldsymbol{\Lambda_b \to \Lambda_c \tau \bar\nu}$ decay}
	\label{sec:B2Lam}
	
	Using the transition form factors in appendix~\ref{sec:form factor}, the helicity amplitudes for the $\Lambda_b \to \Lambda_c$ decay in eq.~(\ref{eq:helicity amplitude}) can be written as
	\begin{align}
	H^{SP}_{\pm 1/2, \pm 1/2}=&F_0g_S \frac{\sqrt{Q_+}}{m_b-m_c} m_- \mp G_0g_P\frac{\sqrt{Q_-}}{m_b+m_c} m_+,
	\nn\\
	H^{VA}_{\pm 1/2, \pm 1/2,t} = &F_0(1+g_L+g_R)\frac{\sqrt{Q_+}}{\sqrt{q^2}} m_- \mp G_0(1+g_L-g_R)\frac{\sqrt{Q_-}}{\sqrt{q^2}} m_+ , 
	\nn\\
	H^{VA}_{\pm 1/2, \pm 1/2,0} = &F_+ (1+g_L+g_R)\frac{\sqrt{Q_-}}{\sqrt{q^2}} m_+  \mp G_+ (1+g_L-g_R)\frac{\sqrt{Q_+}}{\sqrt{q^2}} m_-, 
	\nn\\
	H^{VA}_{\mp 1/2,\pm 1/2,\pm}=&F_\perp (1+g_L+g_R)\sqrt{2Q_-} \mp G_\perp (1+g_L-g_R)\sqrt{2Q_+} , 
	\nn\\
	H^{T,\pm 1/2}_{\pm 1/2,t,0} =&H^{T,\pm 1/2}_{\pm 1/2,-,+} = g_T\Big[h_+\sqrt{Q_-}\pm \widetilde{h}_+\sqrt{Q_+}\Big] ,
	\nn\\
	H^{T,\pm 1/2}_{\mp 1/2,t,\mp}=&\mp H^{T,\pm1/2}_{\mp 1/2,0,\mp} =g_T\frac{\sqrt{2}}{\sqrt{q^2}}\Big[h_\perp m_+\sqrt{Q_-}\mp\widetilde{h}_\perp m_- \sqrt{Q_+}\Big] ,
	\end{align}
	with $m_\pm = m_{\Lambda_b}\pm m_{\Lambda_c}$. Then, the differential decay width in eq.~(\ref{eq:dga}) can be written as
	\begin{align}
	\frac{{ \rm d}\Gamma}{{\rm d} q^2} = & N_{\Lambda_c} \biggl[ A_1^{VA}+\frac{m_\tau^2}{2q^2}A_2^{VA} +\frac{3}{2}A_3^{SP}+8\Big(1+\frac{2m_\tau^2}{q^2}\Big)A_4^{T}+\frac{3m_\tau}{\sqrt{q^2}} (A_5^{VA-SP}+ 4 A_6^{VA-T}) \biggr],
	\end{align}
	with
	\begin{align}
	N_{\Lambda_c} & = \frac{G_{F}^{2}|V_{cb}|^2}{384\pi^{3}}\frac{q^2\sqrt{Q_+ Q_-}}{m_{\lb}^3}\Big(1-\frac{m_\tau^2}{q^2}\Big)^2,
	\nn\\
	A_1^{VA}=&|H^{VA}_{-1/2,1/2,+}|^2+ \sum |H^{VA}_{s,s,0}|^2+|H^{VA}_{1/2,-1/2,-}|^2 ,
	\nn\\
	A_2^{VA}=& A_1^{VA}+3 \sum |H^{VA}_{s,s,t}|^2,
	\nn\\
	A_3^{SP}=&\sum |H^{SP}_{s,s}|^2 ,
	\nn\\
	A_4^{T}=&\sum |H^{T,s}_{s,t,0}|^2+|H^{T,1/2}_{-1/2,t,-}|^2+|H^{T,-1/2}_{1/2,t,+}|^2,
	\nn\\
	A_5^{VA-SP}=&\sum \Re[H^{SP*}_{s,s} H^{VA}_{s,s,t}] ,
	\nn\\
	A_6^{VA-T}=&\sum \Re[H^{VA*}_{s,s,0}H^{T,s}_{s,t,0}] + \Re[H^{VA*}_{-1/2,1/2,+}H^{T,-1/2}_{1/2,t,+}]+ \Re[H^{VA*}_{1/2,-1/2,-}H^{T,1/2}_{-1/2,t,-}],
	\end{align}
	where $\sum$ means the summation over $s=\pm 1/2$. For the forward-backward asymmetry in eq.~(\ref{eq:AFB}), we have
	\begin{align}
	\frac{ {\rm d} A_{\rm FB}} { {\rm d} q^2}=\frac{ N_{\Lambda_c} }{{\rm d} \Gamma / {\rm d} q^2}\frac{3}{4} \biggl[ B_1^{VA}+\frac{2m_\tau^2}{q^2} \bigl(B_2^{VA}+8 B_3^T \bigr)+ \frac{2m_\tau}{\sqrt{q^2}} \bigl(B_4^{VA-SP} +4 B_5^{VA-T} \bigr) + 8B_6^{SP-T} \biggr],
	\end{align}
	where
	\begin{align}
	B_1^{VA}=&~|H^{VA}_{-1/2,1/2,+}|^2-|H^{VA}_{1/2,-1/2,-}|^2 ,
	\nn\\
	B_2^{VA}=&~\sum \Re[H^{VA*}_{s,s,t}H^{VA}_{s,s,0}]  ,
	\nn\\
	B_3^{T}=&~|H^{T,-1/2}_{1/2,t,+}|^2-|H^{T,1/2}_{-1/2,t,-}|^2, 
	\nn\\
	B_4^{VA-SP}=& \sum \Re[H^{SP*}_{s,s}H^{VA}_{s,s,0}], 
	\nn\\
	B_5^{VA-T}=& \sum\Re[H^{VA*}_{s,s,t} H^{T,s}_{s,t,0}]+\Re[H^{VA*}_{-1/2,1/2,+} H^{T,-1/2}_{1/2,t,+}]-\Re[H^{VA*}_{1/2,-1/2,-} H^{T,1/2}_{-1/2,t,-}], 
	\nn\\
	B_6^{SP-T}=&~\sum\Re[H^{SP*}_{s,s}H^{T,s}_{s,t,0}].
	\end{align}
	For the $\Lambda_c$ longitudinal polarization fraction in eq.~(\ref{eq:PL}), we have
	\begin{align}
	\frac{{\rm d}P_{L}^{\lc}}{{\rm d} q^2}=&\frac{N_{\Lambda_c}}{{\rm d}\Gamma/{\rm d}q^2}\frac{1}{2}\biggl[2 C_1^{VA}+\frac{m_\tau^2}{q^2}C_2^{VA}+3C_3^{SP}
	\nn\\
	&\hspace{5em}+16 \Bigl(1 + \frac{2m_\tau^2}{q^2} \Bigr)C_4^{T} +6\frac{m_\tau}{\sqrt{q^2}} \bigl(C_5^{VA-SP}+ 4C_6^{VA-T} \bigr) \biggr],
	\end{align}
	where
	\begin{align}
	C_1^{VA}=&|H^{VA}_{1/2,1/2,0}|^2-|H^{VA}_{-1/2,-1/2,0}|^2+|H^{VA}_{-1/2,1/2,+}|^2-|H^{VA}_{1/2,-1/2,-}|^2,
	\nn\\
	C_2^{VA}=&C_1^{VA} -3|H^{VA}_{-1/2,-1/2,t}|^2 +3|H^{VA}_{1/2,1/2,t}|^2, 
	\nn\\
	C_3^{SP}=&|H^{SP}_{1/2,1/2}|^2-|H^{SP}_{-1/2,-1/2}|^2, 
	\nn\\
	C_4^{T}=&\sum 2s|H^{T,s}_{s,t,0}|^2+|H^{T,-1/2}_{1/2,t,+}|^2 -|H^{T,1/2}_{-1/2,t,-}|^2, 
	\nn\\
	C_5^{VA-SP}=& \sum 2s\Re[H^{SP*}_{s,s}H^{VA}_{s,s,t}], 
	\nn\\
	C_6^{VA-T}=&\sum 2s\Re\big[H^{T,s}_{s,t,0}H^{VA*}_{s,s,0}\big]+\Re\big[H^{T,-1/2}_{1/2,t,+}H^{VA*}_{-1/2,1/2,+}\big] -\Re\big[H^{T,1/2}_{-1/2,t,-}H^{VA*}_{1/2,-1/2,-}\big].
	\end{align}
	For the $\tau$-lepton longitudinal polarization fraction, we have
	\begin{align}
	\frac{{\rm d} P_L^\tau}{{\rm d} q^2}=&\frac{N_{\Lambda_c}}{{\rm d}\Gamma/{\rm d}q^2}\frac{1}{2}\biggl[-2 D_1^{VA}+\frac{m_\tau^2}{q^2}D_2^{VA}+3D_3^{SP} \nn\\
	&\hspace{5em}+16 \Bigl(1-\frac{2m_\tau^2}{q^2} \Bigr)D_4^{T}+\frac{m_\tau}{\sqrt{q^2}} \bigl( 6D_5^{VA-SP}-8D_6^{VA-T} \bigr) \biggr],
	\end{align}
	where
	\begin{align}
	D_1^{VA}=& \sum |H^{VA}_{s,s,0}|^2 + |H^{VA}_{-1/2,1/2,+}|^2 + |H^{VA}_{1/2,-1/2,-}|^2,
	\nn\\
	D_2^{VA}=& D_1^{VA}+3\sum |H^{VA}_{s,s,t}|^2, 
	\nn\\
	D_3^{SP}=& \sum |H^{SP}_{s,s}|^2, 
	\nn\\
	D_4^{T}=& \sum |H^{T,s}_{s,t,0}|^2+|H^{T,-1/2}_{1/2,t,+}|^2 + |H^{T,1/2}_{-1/2,t,-}|^2 ,
	\nn\\
	D_5^{VA-SP}=& \sum \Re[H^{SP*}_{s, s}H^{VA}_{ s, s,t}], 
	\nn\\
	D_6^{VA-T}=&\sum \Re[H^{T,s}_{s,t,0}H^{VA*}_{s,s,0}]
	+\Re[H^{T,-1/2}_{1/2,t,+}H^{VA*}_{-1/2,1/2,+}]+\Re[H^{T,1/2}_{-1/2,t,-}H^{VA*}_{1/2,-1/2,-}].
	\end{align}

\end{appendix}

%\bibliographystyle{JHEP}
%\bibliography{ref}

\providecommand{\href}[2]{#2}\begingroup\raggedright\endgroup

\end{document}